\documentclass[%
nofootinbib,
 amsmath,amssymb,
 aps,
 prd,
]{revtex4-2}

\usepackage{slashed}
\usepackage{graphicx}
\usepackage{dcolumn}
\usepackage{bm}
\usepackage{hyperref}

\usepackage{float}

\newcommand{\forests}{\mathfrak{F}}
\newcommand{\infragr}{\mathcal{I}}

\newcommand{\ww}{\xi}
\newcommand{\cckk}{\mathcal{K}}
\newcommand{\ssbr}{\mathcal{S}}

\begin{document}

\title{
Calculation of lepton magnetic moments in quantum electrodynamics: a justification of the flexible divergence elimination method
}

\author{Sergey Volkov}%
 \email{sergey.volkov.1811@gmail.com, volkoff_sergey@mail.ru, sergey.volkov@partner.kit.edu}
\affiliation{%
 Institute for Theoretical Physics, Karlsruhe Institute of Technology (KIT), \\
76128 Karlsruhe, Germany
}%


\begin{abstract}
The flexible method of reduction to finite integrals,  briefly described in earlier publications of the author, is described in detail. The method is suitable for the calculation of all quantum electrodynamical contributions to the magnetic moments of leptons. It includes mass-dependent contributions. 

The method removes all divergences (UV, IR and mixed) point-by-point in Feynman parametric space without any usage of limit-like regularizations. It yields a finite integral for each individual Feynman graph. The subtraction procedure is based on the use of linear operators applied to the Feynman amplitudes of UV-divergent subgraphs; a placement of all terms in the same Feynman parametric space is implied. The final result is simply the sum of the individual graph contributions; no residual renormalization is required. The method also allows us to split the total contribution into the contributions of small gauge-invariant classes.

The procedure offers a great freedom in the choice of the linear operators. This freedom can be used for improving the computation speed and for a reliability check. 

The mechanism of divergence elimination is explained, as well as the equivalence of the method and the on-shell renormalization. 

For illustrative purposes, all 4-loop contributions to the anomalous magnetic moments of the electron and muon are given for each small gauge-invariant class, as well as their comparison with previously known results. This also includes the contributions that depend on the ratios of the tau-lepton mass to the electron and muon mass.
\end{abstract}

\maketitle


\section{Introduction}

\subsection{Quantum field theory, divergences and renormalization}


It is well known that the first formulations of quantum field theory suffered from ultraviolet (UV) divergences. In 1947-1949 the first results appeared showing that the UV divergences in the physical observables are cancelled if we define the physical parameters properly. This demonstration was the result of the efforts of R. Feynman, J. Schwinger, H. Bethe, S. Tomonaga and others. A procedure for dealing with UV divergences in quantum electrodynamics (QED) at any order of the perturbation series was developed by F. J. Dyson \cite{dyson} and A. Salam \cite{salam}. A further development and a mathematical understanding of these ideas has led to N. Bogoliubov's $\mathcal{R}$-operation. This operation can be represented as a subtraction procedure that deals with the Feynman amplitudes of the UV-divergent subgraphs in each Feynman graph that contributes to the needed probability amplitude; it removes all UV divergences in each Feynman graph. The application of the $\mathcal{R}$-operation to Feynman graphs is equivalent to an introduction of counterterms into the Lagrangian; it can be treated as a renormalization of the theory parameters. The corresponding theorem was proved in 1956 by N. Bogoliubov and O. Parasiuk \cite{bogolubovparasuk}. This proof contains some inaccuracies, but nevertheless, the Bogoliubov-Parasiuk theorem set a standard of quality and raised the hope that quantum field theory has a meaning and can be rigorously examined. The shortcomings of the proof were corrected by K. Hepp in 1964 \cite{hepp}.

Not only the Bogoliubov-Parasiuk-Hepp theorem is of interest, but also its proof, because a stronger statement was proved: The UV subtractions lead directly to finite integrals in Schwinger parametric space\footnote{Usually, these statements are referred to as the BPHZ approach. The last letter ``Z'' refers to W. Zimmermann's paper \cite{zimmerman}. In that work, the convergence theorem was proved directly in Minkowsky momentum space. Moreover, an explicit solution of the recurrence relations given by the $\mathcal{R}$-operation was provided, also known as Zimmermann's forest formula (more precisely, the solution had been provided four years earlier by O. I. Zavialov and B. M. Stepanov \cite{zavialovstepanov}, but this publication was only in Russian and was not noticed).}. In QED we have Feynman graphs with the propagators 
\begin{equation}\label{eq_propagators} 
\frac{i(\slashed{q}+m)}{q^2-m^2+i\varepsilon},\quad \frac{-g_{\mu\nu}}{q^2+i\varepsilon}
\end{equation} 
for the lepton and photon lines correspondingly, where $\slashed{q}=q^{\nu}\gamma_{\nu}$ stands for the Dirac gamma matrices $\gamma_{\nu}$. If we transfer to the Schwinger parameters by using the formula
$$
\frac{i}{x}=\int_0^{+\infty} e^{ixz} dz,
$$
we express the Feynman amplitude as an integral
\begin{equation}\label{eq_schwinger_parametric}
\int_{z_1,\ldots,z_L>0} I(z_1,\ldots,z_L,p_1,\ldots,p_n,\varepsilon) dz_1\ldots dz_L,
\end{equation}
where $p=(p_1,\ldots,p_n)$ is the vector of external momenta, $z=(z_1,\ldots,z_L)$ is the vector of Schwinger parameters, $z_j$ corresponds to the $j$-th internal line of the graph. If we apply the $\mathcal{R}$-operation, $I(z_1,\ldots,z_L,p_1,\ldots,p_n,\varepsilon)$ is expressed as a linear combination of terms; each term is a product of some functions, each of which corresponds to a Feynman amplitude that uses a subset of the set of internal lines; if the same mapping is used between $z_j$ and the internal lines, all the counterterms are placed in the same Schwinger parametric space. A byproduct of the Bogoliubov-Parasiuk-Hepp theorem is that (\ref{eq_schwinger_parametric}) is finite for any $\varepsilon>0$.

Unfortunately, these subtractions in Schwinger's parametric space are useless for calculations. The reason is the inability to handle the limit $\varepsilon\rightarrow 0$. Infrared (IR) divergences and their cancellation have been widely discussed in literature. However, a rigorous examination for the general case has not yet been done. At higher orders IR and UV divergences mix with each other\footnote{See, e.g., the explicit formulas in \cite{adkins}}. In the case of magnetic moments of leptons, all IR divergences are removed by the physical on-shell renormalization as well as the UV and mixed ones\footnote{See, for example, \cite{cvitanovic_gauge} and Appendix C of \cite{grozin}.}. A general scaterring process requires consideration of a finite photon detector sensitivity in addition to renormalization. The on-shell renormalization can be performed in place in Feynman graphs by a simple modification of the $\mathcal{R}$-operation. In the case of the lepton magnetic moments, it removes all divergences in the final result. However, the individual integrals remain IR divergent.

The combinatorics of the IR and mixed divergences in Feynman graphs is very complicated. For this reason, their subtraction before integration is very rarely used in modern quantum field theory computations. Nowadays, most calculations use dimensional regularization, which allows us to work directly with infinite components without exact mappings for cancellation. However, the use of dimensional regularization requires an enormous amount of symbolic manipulation at higher orders. An experience with calculations of the lepton magnetic moment shows that the methods that substract divergences before integration work much faster and allow us to obtain high-order corrections that are not achievable with other methods.

\subsection{Anomalous magnetic moments of leptons and their calculations}

The anomalous magnetic moments (AMM) of the electron and muon are known with a very high precision. A recent measurement \cite{experiment_electron_2022} gave the result
\begin{equation}\label{eq_experiment}
a_e=0.00115965218059(13)
\end{equation}

Standard Model predictions for the electron AMM $a_e$ use the following expression:
$$
a_e=a_e(\text{QED})+a_e(\text{hadronic})+a_e(\text{electroweak}),
$$
$$
a_e(\text{QED})=\sum_{n\geq 1} \left(\frac{\alpha}{\pi}\right)^n
a_e^{2n},
$$
$$
a_e^{2n}=A_1^{(2n)}+A_2^{(2n)}(m_e/m_{\mu})+A_2^{(2n)}(m_e/m_{\tau})+A_3^{(2n)}(m_e/m_{\mu},m_e/m_{\tau}),
$$
where $m_e,m_{\mu},m_{\tau}$ are the masses of the electron, muon, and tau-lepton, respectively, $\alpha$ is the fine-structure constant; a similar expression is used for the muon AMM $a_{\mu}$. 

The universal QED terms $A_1^{(2n)}(\alpha/\pi)^n$ form the main contribution to $a_e$ and $a_{\mu}$. The value
$$
A_1^{(2)}=0.5
$$
was obtained in 1948 by J. Schwinger \cite{schwinger1,schwinger2}. The 2-loop contribution $A_1^{(4)}$ was calculated mainly by R. Karplus and N. Kroll \cite{karpluskroll}. However, this calculation had an error; the correct value
$$
A_1^{(4)}=-0.328478965579\ldots
$$
was independently presented by A. Petermann \cite{analyt2_p} and C. Sommerfield \cite{analyt2_z} in 1957. The value of $A_1^{(6)}$ was being calculated in 1970-x by various research  groups using numerical integration (\cite{carrollyao,carroll}; \cite{levinewright}; \cite{kinoshita_6}); each of these groups used its own method of divergence elimination at the integrand level; the most accurate value $A_1^{(6)}=1.195\pm 0.026$ for those times was obtained in 1974 by T. Kinoshita and P. Cvitanovi\'{c}; the uncertainty is caused by the statistical error of the Monte Carlo integration. At the same time, an analytical calculation of $A_1^{(6)}$ using computers was started. The final value 
$$
A_1^{(6)}=1.181241456\ldots
$$
was obtained by S. Laporta and E. Remiddi in 1996 \cite{analyt3}. This value was the result of the efforts of many researchers (see, for example, 
\cite{analyt_mi, analyt_b3, analyt_j, analyt_e, analyt_d, laporta_1993}
). First numerical estimates for $A_1^{(8)}$ were obtained by T. Kinoshita and W. B. Lindquist in 1981 \cite{kinoshita_8_first}. The most accurate value presented by T. Kinoshita's team, $A_1^{(8)}=-1.91298(84)$, was published in 2015 \cite{kinoshita_8_last}. This value was obtained by Monte Carlo integration. The semianalytic result of S. Laporta 
$$
A_1^{(8)}=-1.9122457649\ldots
$$
was obtained in 2017 \cite{laporta_8}. These two calculations of $A_1^{(8)}$ agree well, as do another independent calculations ~\cite{smirnov_amm,rappl} and our calculation of Feynman graphs without lepton loops \cite{volkov_gpu}. First overall calculation results for $A_1^{(10)}$ were published in 2012 by T. Aoyama, M. Hayakawa, T. Kinoshita, M. Nio in \cite{kinoshita_10_first}\footnote{And another subtraction procedure that removes all divergences before integration was developed for these calculations; it is different from those used for 6th and 8th order calculations.}. The last value obtained by this group in 2019 \cite{kinoshita_atoms} is
\begin{equation}\label{eq_kinoshita_10}
A_1^{(10)}[\text{AHKN}]=6.737(159).
\end{equation}
This coefficient has not yet been verified, and a significant computational error could be apparent in experiments. We recalculated the total contribution of the graphs without lepton loops to $A_1^{(10)}$ and presented \cite{volkov_prd} in 2019 the value 
$$
A_1^{(10)}[\text{no lepton loops,Volkov}]=6.793(90)
$$
that leads to 
\begin{equation}\label{eq_volkov_10}
A_1^{(10)}[\text{Volkov+AHKN}]=5.862(90)
\end{equation}
with a discrepancy of $4.8\sigma$ with (\ref{eq_kinoshita_10}). At the moment the discrepancy is unresolved, but independent calculations are coming \cite{kitanotakaura}. The values (\ref{eq_kinoshita_10}) and (\ref{eq_volkov_10}) in combination with the experimental value (\ref{eq_experiment}) and another known contributions \cite{kinoshita_atoms} lead to
\begin{equation}\label{eq_alpha_kinoshita}
\alpha^{-1}[a_e,\text{AHKN}]=137.0359991663(155)
\end{equation}
and
\begin{equation}\label{eq_alpha_volkov}
\alpha^{-1}[a_e,\text{Volkov+AHKN}]=137.0359991593(155),
\end{equation}
correspondingly. The values obtained from the measured ratios of the atomic masses and the Planck constant
$$
\alpha^{-1}[\text{Rb-2011}]=137.035998996(85), \quad \alpha^{-1}[\text{Cs-2018}]=137.035999046(27),\quad \alpha^{-1}[\text{Rb-2020}]=137.035999206(11)
$$
come from \cite{alpha_rubidium}, \cite{alpha_cesium}, and \cite{alpha_rubidium_2020}, respectively. Note that $\alpha^{-1}[\text{Rb-2020}]$ is the largest among these three values and has  a discrepancy of $5.4\sigma$ relative to $\alpha^{-1}[\text{Cs-2018}]$. The tensions with (\ref{eq_alpha_kinoshita}) are $1.97\sigma$, $3.86\sigma$, $2.09\sigma$; the corresponding tensions with (\ref{eq_alpha_volkov}) are $1.89\sigma$, $3.64\sigma$, $2.46\sigma$.

Other results exist for small classes of 5-loop and higher order graphs \cite{lautrup_bubble,ladders_12,muon_10_baikov}; they are in good agreement with the ones mentioned above.

An analytic formula for $A_2^{(4)}(x)$ was first established by H. H. Elend \cite{elend} in 1966. Numerical values for $A_2^{(6)}$ appeared about 1969 \cite{muon_6_a2_ll_kinoshita}. Most attention was paid to $A_2^{(6)}(m_{\mu}/m_e)$, because its value $A_2^{(6)}(m_{\mu}/m_e)\approx 20$ became unexpectedly large. These values were also being calculated analytically in the form of an expansion in the mass ratio. These expansions are being published since 1975 with an increasing number of terms \cite{a2_6_vacpol_barbieri_remiddi, a23_6_li_samuel, a2_6_vacpol_laporta, a2_6_ll_laporta_remiddi, a2_6_passera}. The values $A_3^{(6)}$ were also evaluated numerically \cite{muon_8_kinoshita_first}, semianalytically \cite{a23_6_li_samuel} and analytically \cite{muon_6_a3_czar} as expansions in the mass ratios. First numerical values of $A_2^{(8)}(m_{\mu}/m_e)$ appeared in 1990 \cite{muon_8_kinoshita_first}. The recent values of $A_2^{(8)}$ and $A_3^{(8)}$ for the electron and muon obtained by T. Aoyama, M. Hayakawa, T. Kinoshita, M. Nio with the help of Monte Carlo integration are presented in \cite{kinoshita_8_muon,kinoshita_8_last}. Semianalytic calculations using expansions in mass ratios confirmed these results \cite{kurz_8_light,kurz_8_heavy}. Other results for certain classes of graphs are also in good agreement with them \cite{muon_8_vacpol_laporta, muon_8_baikov_broadhurst, aguilar_rafael_8, melnikov_8, kataev_8, solovtsova_2019, solovtsova_2022}. The only known values of $A_2^{(10)}$ and $A_3^{(10)}$ for the electron and muon were presented in 2012 by T. Aoyama, M. Hayakawa. T. Kinoshita, M. Nio \cite{kinoshita_10_first, kinoshita_8_muon}. Partial calculations of small graph classes confirm these results \cite{a23_6_li_samuel, aguilar_rafael_8, muon_12_baikov_chetyrkin, muon_10_baikov}. However, a very important contributing to $a_{\mu}$ value
$$
A_2^{(10)}(m_{\mu}/m_e)=742.18(87)
$$
has not been double-checked yet. It is much larger (in absolute value) than the corresponding electron value. A significant error in this value could be noticeable in experiments, and the shift would be comparable to the hadronic uncertainty. In addition, rapid growth of $A_2^{(2n)}(m_{\mu}/m_e)$ with $n$ may cause the higher-order terms to significantly affect the result. However, estimates based on known lower-order values and renormgroup-inspired arguments \cite{muon_8_kinoshita_first, kataev_10_est_1992, kataev_10_est_1995, kataev_10_est_2006, kinoshita_8_muon}, as well as non-relativistic calculations \cite{karshenboim_10}, show that both possibilities seem unlikely.

See reviews \cite{muon_sm_review} on the muon (2021) and \cite{kinoshita_atoms} on the electron (2019) for details.

\subsection{Methods of numerical calculation and gauge-invariant classes}

Methods based on subtraction of divergences before integration have been widely used for calculations of  QED contributions to magnetic moments of leptons \cite{levinewright, kinoshita_6, carrollyao, kinoshita_atoms} and for other quantum field theory calculations \cite{kompaniets}.

In 2016, we presented a method \cite{volkov_2015} suitable for computing $A_1^{(2n)}$ for arbitrary $n$. The difference of this method from the previously known methods is that it is based on linear operators applied to the Feynman amplitudes of UV-divergent subgraphs, and the final result is obtained directly by summing the contributions of the Feynman graphs (no residual renormalization is required).

A distribution of the final result over Feynman graphs is a matter of free choice, but it is very untrivial to realize this freedom in the context of methods that subtract divergences before integration and yield the final result without residual renormalization, because the integrals obtained must be finite for each individual Feynman graph, but the whole procedure must conform to the physical on-shell renormalization. In a sense, this subtraction procedure must contain an evidence that the on-shell renormalization removes all divergences, including infrared and mixed divergences. In 2021, we presented a modification \cite{volkov_ffk2021} of the developed method; it allows us to choose 3 linear subtraction operators independently\footnote{The number is not related to the number of leptons; the number of leptons can be arbirtarily large.}. Moreover, the new method is suitable for the calculation of mass-dependent contributions\footnote{More precisely, the old method is also suitable for this, but the argument in \cite{volkov_2015} does not prove this and is not extensible to prove this.} $A_2^{(2n)}$, $A_3^{(2n)}$.

Some classes of Feynman graphs form gauge-invariant classes. For example, any class of graphs closed under a motion of the internal photon lines along the lepton paths and loops, but without jumping over the external photon line, is invariant in the class of Lorentz-invariant gauges with the photon propagators of the form $-(g_{\mu\nu}+k_{\mu}k_{\nu}f(k^2))/(k^2+i0)$, if we properly define the on-shell renormalization for subclasses of graphs\footnote{The gauge invariance was proved in \cite{cvitanovic_gauge}, but only for graphs without lepton loops. It looks the whole proof is not published, but the fact is widely used; we believe that the ideas of \cite{cvitanovic_gauge} can be extended to the general case.} \footnote{The gauge invariance implies not only the possibility of using different photon propagators, but also the ability to renormalize photon self-energies differently.}, see Section \ref{subsec_on_shell_definition}.

Examples of the use of the new method, as well as a comparison with the old method, were given in \cite{volkov_ffk2021, volkov_acat_2021}. In this work we show that the method yields the exact value for each gauge-invariant class mentioned above, including those contributing to $A_2^{(2n)}$ and $A_3^{(2n)}$, as well as a finite integral for each Feynman graph.

\section{The method formulation}

We work in the unit system where $\hbar=c=1$, the factors of $4\pi$ appear in the fine-structure constant: $\alpha=e^2/(4\pi)$, the tensor $g_{\mu\nu}$ corresponds to the signature $(+,-,-,-)$, the Dirac matrices satisfy the condition $\gamma_{\mu}\gamma_{\nu}+\gamma_{\nu}\gamma_{\mu}=2g_{\mu\nu}$. We also suppose $e=1$ for the lepton electric charge; since we work with each term of the perturbation series separately, it changes nothing, but makes the expressions shorter.

We extract the lepton AMM from QED Feynman graphs with $N_l=2$, $N_{\gamma}=1$, where by $N_l$ and $N_{\gamma}$ we denote the number of external lepton and photon lines in the graph; each graph may contain electron, muon, tau-lepton lines. We also assume that all graphs are one-particle irreducible and have no odd lepton loops (Furry's theorem).

We work in the Feynman gauge with the propagators (\ref{eq_propagators}) for leptons and photons, respectively, where $m$ is the lepton mass.

There are the following types of UV-divergent subgraphs\footnote{We consider only those subgraphs which are one-particle irreducible and contain all lines connecting the vertexes of the given subgraph; since odd lepton loops are forbidden, a UV-divergent subgraph is one-particle irreducible if and only if it is amputated.} in QED Feynman graphs: \emph{lepton self-energy} subgraphs ($N_l=2$, $N_{\gamma}=0$), \emph{vertexlike} subgraphs ($N_l=2$, $N_{\gamma}=1$), \emph{photon self-energy} subgraphs ($N_l=0$, $N_{\gamma}=2$), \emph{photon-photon scattering} subgraphs\footnote{Photon-photon scattering subgraph divergences cancel in the final result without subtraction, but they remain in the individual graphs.} ($N_l=0$, $N_{\gamma}=4$).

Two subgraphs are said to overlap if they are not contained in each other and the intersection of their line sets is not empty.

A set of subgraphs of a graph is called a \emph{forest} if any two elements of this set do not overlap.

For a vertexlike graph $G$, we denote by $\forests[G]$  the set of all forests $F$ consisting of UV-divergent subgraphs of $G$ and satisfying the condition $G\in F$. The lepton path of a graph $G$ connecting its external lepton lines is called the \emph{main path} of $G$. By $\infragr[G]$ we denote\footnote{The definition differs from that in \cite{volkov_2015}.} the set of all vertexlike subgraphs of $G$ (including $G$) which have the vertex incident to the external photon of $G$ and at least one vertex of the main path of $G$.

To define the subtraction procedure, we use the linear operators labeled $A$, $U_0,U_1,U_2,U_3$, $L$; sometimes the same names are used for operators applied to the Feynman amplitudes of subgraphs of different types. The definitions are:
\begin{itemize}
\item $A$ is applied to vertexlike Feynman amplitudes\footnote{$p-\frac{q}{2}$, $p+\frac{q}{2}$ are incoming and outgoing lepton momenta, $q$ is the photon momentum.} $\Gamma_{\mu}(p,q)$ and is defined as
$$
(A\Gamma)_{\mu}(p,q)=[A'\Gamma]\gamma_{\mu},
$$
where $A'$ is the projector of the AMM. See the definition of the projector in \cite{volkov_2015,volkov_prd}.
\item $L$ is the standard on-shell renormalization operator for vertexlike graphs; it is defined as
$$
(L\Gamma)_{\mu}(p,q)=[L'\Gamma]\gamma_{\mu},\quad L'\Gamma=a(m^2)+mb(m^2)+m^2c(m^2),
$$
where
\begin{equation}\label{eq_vertexlike}
\Gamma_{\mu}(p,0)=a(p^2)\gamma_{\mu} + b(p^2)p_{\mu} + c(p^2) \slashed{p}p_{\mu} + d(p^2)(\slashed{p}\gamma_{\mu}-\gamma_{\mu}\slashed{p}).
\end{equation}
\item $U_0$ is applied to photon self-energy and photon-photon scattering subgraphs and works as in standard renormalization. For the Feynman amplitude $\Pi_{\mu\nu}(p^2)$ of the photon self-energy, we can take, for example,
\begin{equation}\label{eq_u0_pse_def}
(U_0 \Pi)_{\mu\nu}(p^2)=[U_0' \Pi][g_{\mu\nu}, p^2 g_{\mu\nu},p_{\mu}p_{\nu}],
\end{equation}
where
\begin{equation}\label{eq_u0s_pse_def}
U_0' \Pi =\left[ h_1(0), \left. \frac{\partial h_1(p^2)}{\partial p^2} \right|_{p^2=0}, h_2(0) \right],
\end{equation}
a scalar product is implied, and
\begin{equation}\label{eq_pi_def}
\Pi_{\mu\nu}(p^2)=h_1(p^2)g_{\mu\nu}+h_2(p^2) p_{\mu}p_{\nu}.
\end{equation}
Optimizations as described in \cite{kinoshita_infrared} can also be used for photon self-energy graphs; the choice of the photon self-energy renormalization method does not play an important role in our subtraction procedure. For example, a more efficient method was used for the calculations; see Section \ref{sec_computation}. However, the definition given here is convenient for proofs.

For the Feynman amplitudes of photon-photon scattering,  we can take
$$
(U_0 \Pi)_{\mu_1\mu_2\mu_3\mu_4}(p_1,p_2,p_3,p_4)=\Pi_{\mu_1\mu_2\mu_3\mu_4}(0,0,0,0),
$$
where zero 4-momenta are denoted by $0$. It also can be defined with vector-valued operators $U_0'$ as it was done for photon self-energy Feynman amplitudes.
\item $U_j$, $j=1,2$ are applied to vertexlike and lepton self-energy Feynman amplitudes; $U_3$ is applied only to vertexlike amplitudes. They are defined as
$$
(U_j \Gamma)_{\mu}(p,q)=[U_j'\Gamma]\gamma_{\mu},
$$
\begin{equation}\label{eq_u_self_ener}
(U_j\Sigma)(p)=M'\Sigma+(U_j'\Sigma)\times(\slashed{p}-m)
\end{equation}
where $\Gamma_{\mu}(p,q)$ and $\Sigma(p)$ are vertexlike and lepton self-energy Feynman amplitudes,
$U_j'$ are number-valued linear operators,
\begin{equation}\label{eq_sigma}
\Sigma(p)=r(p^2)+s(p^2)\slashed{p}.
\end{equation}
\begin{equation}\label{eq_ms}
M'\Sigma=r(m^2)+s(m^2)m.
\end{equation}

We require that:
\begin{enumerate}
\item $U_j$ ($j=1,2$) \emph{preserve the Ward identity}: if $\Gamma_{\mu}$ and $\Sigma$ satisfy
$$
\Gamma_{\mu}(p,0)=-\frac{\partial \Sigma(p)}{\partial p^{\mu}},
$$
then 
$$
(U_j\Gamma)_{\mu}(p,0)=-\frac{\partial [(U_j\Sigma)(p)]}{\partial p^{\mu}}
$$
is also satisfied. The latter can be rewritten as 
$$
U_j'\Gamma=-U_j'\Sigma.
$$
\item If $\Gamma_{\mu}(p,0)=0$, then $U'_j\Gamma=0$ ($j=1,2,3$). 
\end{enumerate}

We emphasize that the same symbol $U_j$ is used for lepton self-energy and vertexlike Feynman amplitudes; the definitions for the different types are independent from each other, the requirements connect them.

These conditions are sufficient to prove combinatorially that the subtraction procedure defined below is equivalent to on-shell renormalization, provided that all regularization issues are resolved. 

To obtain finite integrals for each graph, more stringent requirements are needed. We take
$$
U_j'\Gamma=a(M^2),\quad U_j'\Sigma=-s(M^2),\quad j=1,2,
$$
where (\ref{eq_vertexlike}), (\ref{eq_sigma}) are satisfied,
$M^2$ is an arbitrary real number (it may be different for different operators).
As $U_3'\Gamma$ we can take $a(M^2)$ or $L'\Gamma$. 

Being an operator means that it is applied to a function and depends only on this function (it does not depend on the internal structure of the corresponding subgraph). However, the same symbol $U_j$ is used for linear operators that are applied to Feynman amplitudes of different types (a type is the set of the external line types); the linear operators under the same symbol $U_j$ for different types can be chosen independently. Of course, $U_j$ can be chosen independently for different $j$. $U_1$ and $U_3$ are defined only for the Feynman amplitudes with the external leptons coinciding with those whose AMM is calculated; $U_2$ are defined for all types of external leptons (for example, it can be chosen independently for the Feynman amplitudes with external electrons, muons and tau-leptons). 
\end{itemize}

The expression for the subtraction and extraction of the AMM corresponding to a Feynman graph $G$ is
$$
\sum_{\substack{F=\{G_1,\ldots,G_n\}\in\forests[G] \\ G'\in\infragr[G]\cap F}} (-1)^{n-1} S^{G'}_{G_1}\ldots S^{G'}_{G_n},
$$
where
$$
S^{G'}_{G''} =
\begin{cases}
A_{G''},\text{ if } G''=G', \\
L_{G''}-(U_1)_{G''}, \text{ if } G''=G\neq G', \\
L_{G''},\text{ if } G' \subset G'' \subset G, \\
(U_{\ww})_{G''},\text{ if } G''\in\infragr[G]\text{ and } G''\subset G', \\
(U_1)_{G''}, \text{ if } G''\notin\infragr[G]\text{ and }G''\text{ has its external leptons on the main path of }G, \\ 
(U_2)_{G''}, \text{ if } G'' \text{ has its external leptons on a lepton loop of }G, \\
(U_0)_{G''}, \text{ if } G'' \text{ is a photon self-energy or photon-photon scattering subgraph};
\end{cases}
$$
here 
\begin{equation}\label{eq_w}
\ww=\begin{cases}3,\text{ if the vertex incident to the external photon of the whole graph }G\text{ lies on a lepton loop of }G, \\
1\text{ otherwise};
\end{cases}
\end{equation}
the index of an operator denotes the subgraph to whose Feynman amplitude it is applied; $G_1\subset G_2$ or $G_1\subseteq G_2$ means $V(G_1)\subset V(G_2)$ or $V(G_1)\subseteq V(G_2)$, where $V(G)$ denotes the set of vertices of $G$. We emphasize that the whole graph $G$ is used in the definition of $\ww$; thus, $\ww$ does not depend on $G'$ and $G''$. At the level of Feynman amplitudes, each term of the expression sequentially transforms the Feynman amplitudes of subgraphs using corresponding linear operators (the subgraphs should be ordered by inclusion in this sequence, from smaller to larger). Finally, the coefficient before $\gamma_{\mu}$ should be taken.

\begin{figure}[H]
	\begin{center}
		\includegraphics[width=60mm]{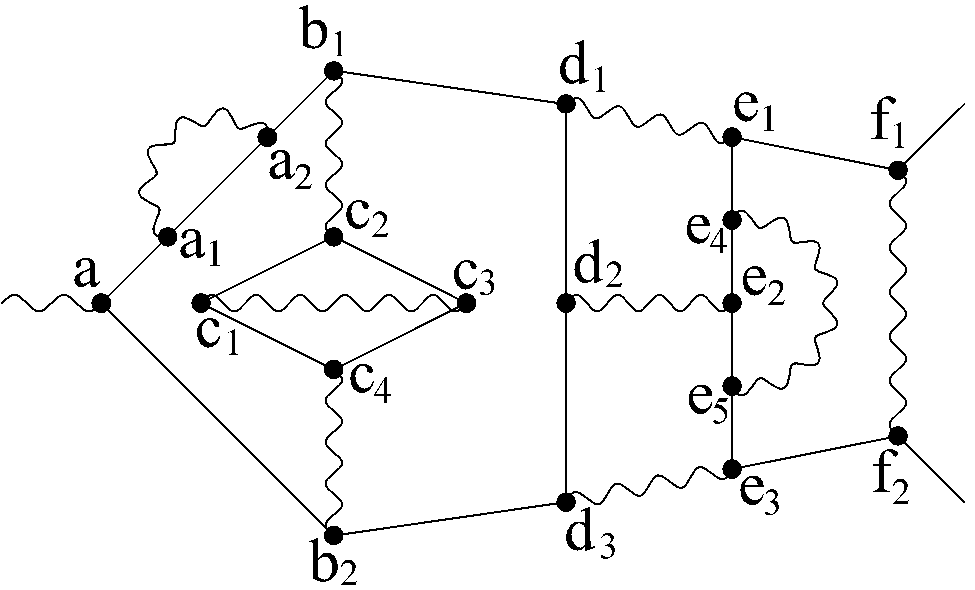}
		\caption{\label{fig_example}An example of a Feynman diagram contributing to AMM.}
	\end{center}
\end{figure}

For example, for the graph $G$ from Fig. \ref{fig_example} the expression is
\begin{gather*}
\left[A_G \left( 1-(U_3)_{G_e} \right) - \left( L_G - (U_1)_G \right) A_{G_e} \right] \times \left( 1-(U_2)_{G_c} \right) \times \left(1 - (U_1)_{e_2 e_4 e_5} \right) \\
\times \left(1-(U_0)_{G_d}\right)\times \left(1-(U_0)_{c_1c_2c_3c_4} \right) \times \left( 1 - (U_2)_{c_1 c_2 c_3} - (U_2)_{c_1 c_3 c_4} \right) \times \left( 1- (U_2)_{a_1a_2} \right),
\end{gather*}
where $G_e=aa_1a_2b_1b_2c_1c_2c_3c_4d_1d_2d_3e_1e_2e_3e_4e_5$, $G_d=\{aa_1a_2b_1b_2c_1c_2c_3c_4d_1d_2d_3\}$, $G_c=aa_1a_2b_1b_2c_1c_2c_3c_4$, $\infragr[G]=\{G_e,G\}$, subdiagrams are denoted by enumeration of their internal vertices; we should expand the parentheses to obtain the forest formula. The expression for this example does not depend on the type of leptons on the lepton loops ($U_2$ may be defined differently for different types of leptons, however).

\begin{figure}[H]
	\begin{center}
		\includegraphics[width=35mm]{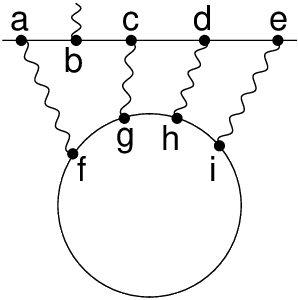}
		\caption{\label{fig_example_iv_d1}Another example of a Feynman diagram contributing to AMM.}
	\end{center}
\end{figure}

Another example is the graph $G$ from Fig. \ref{fig_example_iv_d1}. In this case, $\infragr[G]=\{G\}$, the expression is
$$
A_G \left( 1-(U_1)_{cdefghi} \right) \left( 1 - (U_0)_{fghi} \right).
$$
The operator $U_1$ is applied to the subgraph $cdefghi$, because its external leptons are on the main path of $G$. 

To obtain finite Feynman parametric integrals, all terms should be placed in the same Feynman parametric space. The algorithm for obtaining integrals from expressions is  briefly explained in \cite{volkov_acat_2021} and in detail in \cite{volkov_2015}. The needed value of $A_l^{(2n)}$ is simply the sum of the integrals corresponding to the Feynman graphs involved.

\section{Divergence elimination}

The principle of divergence elimination in the method is similar to that explained in \cite{volkov_2015} for the old method. Here we describe schematically how this choice of linear operators allows us to separate divergences and how it leads to divergence cancellation.

Let us consider the 1-loop unrenormalized vertexlike Feynman amplitude\footnote{We ignore coefficients in this consideration.}
$$
\Gamma_{1,\mu}(p,q) = \int \gamma^{\nu} \frac{\slashed{p}+\frac{\slashed{q}}{2}+\slashed{k}+m}{\left(p+\frac{q}{2}+k\right)^2-m^2+i0} \gamma_{\mu} \frac{\slashed{p}-\frac{\slashed{q}}{2}+\slashed{k}+m}{\left(p-\frac{q}{2}+k\right)^2-m^2+i0} \gamma_{\nu} \frac{1}{k^2+i0}d^4 k.
$$ 
Its IR-divergent part at $p^2=m^2$ is equal to
$$
\Gamma_{1,\mu}(p,0)_{\text{IR}} = \left[\int \gamma^{\nu} \frac{\slashed{p}+m}{\left(p+k\right)^2-m^2+i0} \gamma_{\mu} \frac{\slashed{p}+m}{\left(p+k\right)^2-m^2+i0} \gamma_{\nu} \frac{1}{k^2+i0}d^4 k \right]_{\text{IR}}.
$$
With 
$$
(\slashed{p}+m)\gamma_{\mu}(\slashed{p}+m)=2mp_{\mu}+2p_{\mu}\slashed{p}
$$
and the fact that $p_{\mu}$ commutes with all other multipliers that may appear in the expression, we obtain
$$
\Gamma_{1,\mu}(p,0)_{\text{IR}}=Ap_{\mu}+Bp_{\mu}\slashed{p},
$$
where $A$ and $B$ some IR-divergent values\footnote{We mean that a regularization is introduced, but we omit it for simplicity.}. Thus, $U_1 \Gamma_1$ and $U_2 \Gamma_1$ are IR-finite. On the other hand, the UV-divergent part of $\Gamma_{1,\mu}(p,q)$ does not depend on $p,q$ and is therefore proportional to $\gamma_{\mu}$. Thus, $(L-U_j)\Gamma_1$, $j=1,2$ are UV-finite.

There is a general observation:
\begin{itemize}
\item $(L-U_j)\Gamma$, $j=1,2,3$ has no \emph{overall} UV divergence for any vertexlike Feynman amplitude $\Gamma$ (of any order);
\item $U_j\Gamma$, $j=1,2$ does not have IR divergences at all, provided that the mass renormalization has been done properly (see below);
\item this is also true for $U_j \Sigma$, $j=1,2$, where $\Sigma$ is a lepton self-energy Feynman amplitude. 
\end{itemize}
In particular, this means that $U_1,U_2$ can be used to subtract UV divergences without generating additional IR divergences (unlike $L$).

\begin{figure}[H]
	\begin{center}
		\includegraphics[width=60mm]{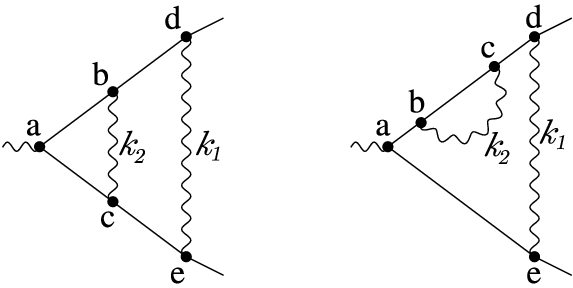}
		\caption{\label{fig_examples_2loop}2-loop examples}
	\end{center}
\end{figure}

Consider the 2-loop ladder graph $G$ in Fig. \ref{fig_examples_2loop},left\footnote{Since changing  the lepton line directions on a path or loop does not change the contribution, we will sometimes work with undirected Feynman graphs.}. The corresponding expression is
$$
A_G - A_G (U_1)_{abc} - (L-U_1)_G A_{abc}.
$$
In this case, we have only logarithmic divergences. The overall UV divergence corresponds to $k_1,k_2\rightarrow\infty$; it is proportional to $\gamma_{\mu}$ and is cancelled in each term by $A$ or $L-U_1$. The overall IR divergence corresponds to $k_1,k_2\rightarrow 0$. In all terms it is cancelled by $A$ since it is proportional to the Born amplitude $(A\gamma)_{\mu}=0$, where by $\gamma$ we denote the 4-vector of Dirac gamma matrices; this can be shown using standard methods for studying IR divergences in QED. There is also a mixed UV-IR divergence in this graph: $k_1\rightarrow 0,k_2\rightarrow\infty$; in this case, the UV divergence corresponding to $abc$ is IR-infinitely enhanced by the remaining propagators. In the first 2 terms it is cancelled by $A$, because the UV-divergent part of $abc$ is proportional to $\gamma_{\mu}$, so the ``enhanced'' value is proportional to $(A\gamma)_{\mu}=0$. In the remaining term, it is cancelled in $A_{abc}$ for the same reason. The remaining divergences\footnote{This investigation is intended to demonstrate ideas only; a rigorous proof is not the goal in this case.} are the UV divergence $k_2\rightarrow \infty$ and the IR divergence $k_1\rightarrow 0$. The first is cancelled in the sum of two terms by $1-U_1$ applied to the amplitude of $abc$ and in the last term by $A_{abc}$. The second does not exist in $A_G (U_1)_{abc}$, in $A_G$ it enhances the amplitude generated by $A_{abc}$ by an IR-infinite multiplier, and it is cancelled by $(L-U_1)_G A_{abc}$, because $L-U_1$ completely extracts the IR-divergent part of the enhancing multiplier.

Let us examine the graph $G$ on Fig. \ref{fig_examples_2loop},right. The expression is
$$
A_G-A_G (U_1)_{bc}.
$$ 
The difficulty in this case is the power-type IR-divergence corresponding to $k_1\rightarrow 0$ and a fixed $k_2\neq 0$. However, it is eliminated by the mass subtraction part in $U_1$. To make this possible, we give no freedom to the definition (\ref{eq_ms}) of $M'$ in (\ref{eq_u_self_ener}). After the mass subtraction, the IR divergence becomes logarithmic and is removed by $A$, as is the overall UV divergence. The UV subdivergence of $bc$ is removed by $1-U_1$, and no additional IR divergence is generated. 

\begin{figure}[H]
	\begin{center}
		\includegraphics[width=40mm]{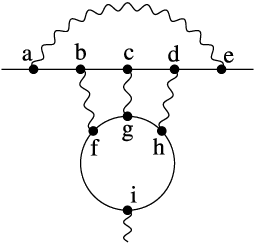}
		\caption{\label{fig_example_with_u3}An example with $U_3$ in the expression}
	\end{center}
\end{figure}

Let us consider the graph $G$ on Fig. \ref{fig_example_with_u3} . The expression is 
$$
\left[A_G - A_G (U_3)_{bcdfghi} - (L-U_1)_G A_{bcdfghi}\right] \left( 1- (U_0)_{fghi} \right).
$$
The last multiplier eliminates the UV divergence corresponding to $fghi$; in this case there are no problems with IR divergences. The last term of the first multiplier removes the IR divergence corresponding to $k_{ae}\rightarrow 0$ (the subscript denotes the graph line by its ends). The term $A_G (U_3)_{bcdfghi}$ removes the UV divergence corresponding to $bcdfghi$. It is important to note that it does not create any additional IR divergence, although $U_3$ has more freedom in definition than $U_1$ and $U_2$ (we can take $U_3=L$, for example). The reason is that the on-shell Feynman amplitude of $bcdfghi$ has no IR divergence, because its external photon is on a lepton loop.

The cancellation of divergences in Feynman parametric space for an arbitrary graph $G$ can be shown in a similar way as in \cite{volkov_2015}, but $U$ should be replaced everywhere by the corresponding $U_j$. Let us note one more small change. Following \cite{volkov_2015}, Appendix C, we assume that $G_1\subset\ldots\subset G_n\subset G_{n+1}=G$ are all elements of $\infragr[G]$. If $\ww=1$, where (\ref{eq_w}) is implied, we also assume that $G_0$ is the graph consisting of only one vertex of $G$ joining the external photon (if $\ww=3$, we use $G_0$ as the empty graph). By $z_l$ we denote the Feynman parameter corresponding to the internal line $l$. With the help of power counting, we investigate the case
$$
z_l\asymp \delta^{\beta_l},\quad \delta\rightarrow 0,
$$
where $\beta_l\geq 0$ are some numbers corresponding to the internal lines. By $P_j$ we denote\footnote{In \cite{volkov_2015} it is denoted by $P_j'\cup P_j''$.} the set of all lines $l$ on the main path of $G$ such that $l\in G_{j+1}$ and $l\notin G_j$. The cancellation of UV divergences and UV parts of mixed divergences is demonstrated in the same way\footnote{The expression is organized in a very similar way to Zimmermann's forest formula; this makes it possible to handle UV divergences.} as in \cite{volkov_2015}. If all divergences not belonging to $\infragr[G]$ are properly subtracted, a potentially IR divergent case corresponds to a \emph{vector of divergences} $[v_0,\ldots,v_n;w_0,\ldots,w_n]$, where $v_j\geq w_j\geq 0$, $2v_0-w_0=2v_1-w_1=\ldots=2v_n-w_n>0$, there exists $j$ such that $w_j=0$. This vector gives the values
$$
\beta_l = \begin{cases} 
v_j,\text{ if } l\in P_j, \\
w_j,\text{ if } l\in G_{j+1},\ j\notin G_j,\ j\notin P_j.
\end{cases}
$$
We will refer to the index $j$, $1\leq j\leq n$ as a \emph{transition index} for the divergence vector $[v,w]$ if $v_j>v_{j-1}$. Correspondingly, the index $j$ is called an \emph{inverse transition index} if $v_j<v_{j-1}$. The difference with the investigation in \cite{volkov_2015} is the following:
\begin{enumerate}
\item An index $j$ is called \emph{separating} in a term of the expression if there is an operator $A_{G_j}$, $L_{G_j}$ or $(U_3)_{G_j}$ ($U_3=L$) in this term. The difference with \cite{volkov_2015} is the possibility to take $U_3$ (if it is equal to $L$).
\item If $\ww=3$, then no consideration of the case $v_0>w_0$ is necessary; this case cannot produce divergence in a term, because $G_1$ has its external photon on a lepton loop.
\item If $\ww=3$ and $U_3=L$, we rewrite the expression differently for a given index $H$ of the inverse transition in $[v,w]$:
$$
-\sum_{1\leq l<H} \left (L_G-(U_1)_G)(1-L_{G_n})\ldots (1-L_{G_{l+1}}) A_{G_l} (1-L_{G_{l-1}})\ldots (1-L_{G_1}) \right) 
$$
$$
+ \sum_{H<l\leq n} \left( (L_G-(U_1)_G) (1-L_{G_n}) \ldots (1-L_{G_{l+1}}) [L_{G_l}A_{G_H}-A_{G_l}] (1-L_{G_{H-1}})\ldots (1-L_{G_1}) \right)
$$
$$
+ [A_G-(L_G-(U_1)_G)A_{G_H}](1-L_{G_{H-1}})\ldots (1-L_{G_1})
$$
$$
+ \sum_{H<l\leq n} \left[ \left( A_G(1-L_{G_n})\ldots (1-L_{G_{l+1}}) - (L_G-(U_1)_G) \sum_{l<k\leq n} \left( A_{G_k} \prod_{\substack{l<r\leq n \\ r\neq k}} (1-L_{G_r}) \right) \right)  \right.
$$
$$
\times 
\left. \vphantom{\prod_{\substack{l<r\leq n \\ r\neq k}}\left[ \left( A_G(1-L_{G_n}) \right) \right] }
 [L_{G_l}L_{G_H}-L_{G_l}](1-L_{G_{H-1}})\ldots (1-L_{G_1})\right].
$$
The last term was added to (C.7) in \cite{volkov_2015}, inappropriate $(1-U)$ multipliers were removed. The factor $L_{G_l}L_{G_H}-L_{G_l}$ cancels the divergence in this term, analogous to the other terms.
\end{enumerate}
This argument shows how it works, but a full rigorous examination seems difficult and cumbersome; we rely on a numerical check (see Section \ref{sec_computation}). 

\section{Equivalence to on-shell renormalization}\label{sec_on_shell}

\subsection{Definition}\label{subsec_on_shell_definition}

The definition of on-shell renormalization is well known for the coefficients of the expansion in $\alpha$, but it requires a definition for subsets of Feynman graphs. We define it as the sum of the individual graph contributions obtained by the in-place on-shell renormalization. It can be written in terms of expressions such as those used above. The expression for a graph $G$ is
$$
\sum_{F=\{G,G_1,\ldots,G_n\}\in\forests[G]} (-1)^n A_G S_{G_1} \ldots S_{G_n},
$$
where
$$
S_{G'}= \begin{cases}
L_{G'},\text{ if }G'\text{ is a vertexlike graph}, \\
B_{G'},\text{ if }G'\text{ is a lepton self-energy graph}, \\
(U_0)_{G'},\text{ if }G'\text{ is a photon self-energy or photon-photon scattering graph};
\end{cases}
$$
here
$$
(B\Sigma)(p)=[B'\Sigma]\times(\slashed{p}-m)+M'\Sigma,\quad B' \Sigma=s(m^2) + 2m\left.\frac{\partial r(x)}{\partial x}\right|_{x=m^2} + 2m^2\left.\frac{\partial s(x)}{\partial x}\right|_{x=m^2},
$$
where (\ref{eq_sigma}) and (\ref{eq_ms}) are satisfied.

The multipliers $S_{G'}$ for photon-photon scattering graphs $G'$ can be omitted in gauge-invariant classes as well as transformations like described in \cite{kinoshita_infrared} can be applied for photon self-energy graphs. 

\subsection{Examples}\label{subsec_equivalence_examples}

2-loop examples are given in \cite{volkov_2015} for the old version of the method. In the new version each operator $U$ should be replaced by $U_1$ or $U_0$. 

\begin{figure}[H]
	\begin{center}
		\includegraphics[width=100mm]{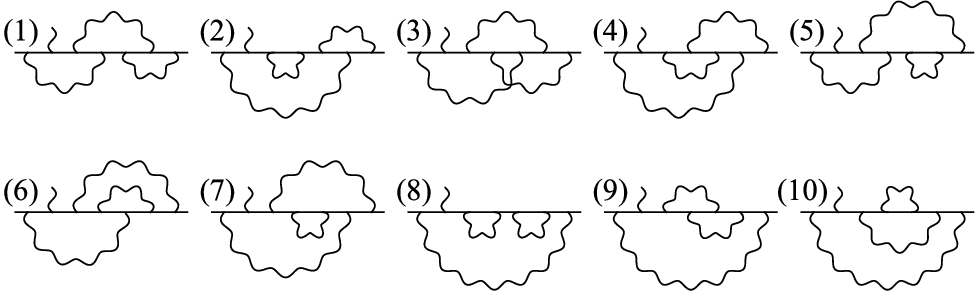}
		\caption{\label{fig_3loops_set3}An example of a 3-loop gauge-invariant class}
	\end{center}
\end{figure}

The Ward identities for individual Feynman graphs play an important role in proving the equivalence. We will illustrate this by the gauge-invariant class shown in Fig. \ref{fig_3loops_set3}. The investigation applies to both the new and old methods (the difference is only in the name of the $U$-operator).

\begin{figure}[H]
	\begin{center}
		\includegraphics[width=100mm]{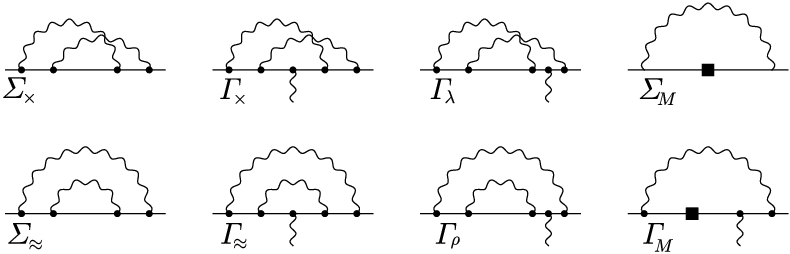}
		\caption{\label{fig_set3_expr_amplitudes}Feynman amplitudes participating in the expressions for the set in Fig. \ref{fig_3loops_set3} and the corresponding Feynman graphs}
	\end{center}
\end{figure}

Table \ref{table_set3_expressions} provides the parts of the contributions, obtained with our method, that contain $U$-operators. The lower-order Feynman amplitudes used in these expressions are shown in Fig. \ref{fig_set3_expr_amplitudes}. The boxed dots represent the mass vertices (the vertices that give $1$ in the Feynman rules). By $\Gamma_1$ and $\Sigma_1$ we also denote the 1-loop vertexlike and lepton self-energy amplitudes (without special vertices), respectively. In this case, $\infragr[G]=\emptyset$, and we have no IR subtractions; therefore, the remaining contributions are the same as those obtained by the in-place on-shell renormalization.

\begin{table}[h]
\caption{\label{table_set3_expressions}Contributions with $U$-operators of the graphs from Fig. \ref{fig_3loops_set3} obtained with our method}
\begin{ruledtabular}
\begin{tabular}{ll}
Set number & Contribution\\
\colrule 
1 & $-(A'\Gamma_1)(U_1'\Gamma_{\lambda}) - (A'\Gamma_{\lambda})(U_1'\Gamma_1) + (A'\Gamma_1)(U_1' \Gamma_1)^2$ \\
2 & $-(A'\Gamma_{\rho})(U_1'\Gamma_1) - (A'\Gamma_{\lambda})(U_1'\Sigma_1) + (A'\Gamma_1)(U_1'\Gamma_1)(U_1'\Sigma_1) + (A'\Gamma_M)(U_1' \Gamma_1)(M'\Sigma_1)$ \\
3 & $-(A'\Gamma_1)(U_1' \Gamma_{\times})$ \\
4 & $-(A'\Gamma_1)(U_1'\Gamma_{\lambda}) - (A'\Gamma_{\lambda})(U_1'\Gamma_1) + (A'\Gamma_1)(U_1'\Gamma_1)^2$ \\
5 & $-(A'\Gamma_1)(U_1'\Gamma_{\rho}) - (A'\Gamma_{\lambda})(U_1'\Sigma_1) + (A'\Gamma_1)(U_1'\Gamma_1)(U_1'\Sigma_1) + (A'\Gamma_1)(U_1'\Gamma_M)(M'\Sigma_1)$ \\
6 & $-(A'\Gamma_1)(U_1'\Gamma_{\approx}) - (A'\Gamma_{\lambda})(U_1'\Gamma_1) + (A'\Gamma_1)(U_1'\Gamma_1)^2$ \\
7 & $-(A'\Gamma_1)(U_1'\Gamma_{\rho}) - (A'\Gamma_{\lambda})(U_1'\Sigma_1) + (A'\Gamma_1)(U_1'\Gamma_1)(U_1'\Sigma_1) + (A'\Gamma_1)(U_1'\Gamma_M)(M'\Sigma_1)$ \\
8 & $-2(A'\Gamma_{\rho})(U_1'\Sigma_1) + (A'\Gamma_1)(U_1'\Sigma_1)^2 +2(A'\Gamma_M)(U_1'\Sigma_1)(M'\Sigma_1)$ \\
9 & $-(A'\Gamma_1)(U_1'\Sigma_{\times}) - 2(A'\Gamma_{\rho})(U_1'\Gamma_1) + 2(A'\Gamma_1)(U_1'\Sigma_1)(U_1'\Gamma_1) + 2(A'\Gamma_M)(M'\Sigma_1)(U_1'\Gamma_1)$ \\
10 & $-(A'\Gamma_1)(U_1'\Sigma_{\approx}) - (A'\Gamma_{\rho})(U_1'\Sigma_1) + (A'\Gamma_1)(U_1'\Sigma_1)^2 + (A'\Gamma_M)(M'\Sigma_1)(U_1'\Sigma_1) + (A'\Gamma_1)(U_1'\Sigma_M)(M'\Sigma_1)$
\end{tabular}
\end{ruledtabular}
\end{table}

After summation, we obtain
$$
(A' \Gamma_1)(-[2U_1'\Gamma_{\lambda} + U_1' \Gamma_{\times} + U_1' \Sigma_{\times}] - [2U_1'\Gamma_{\rho} + U_1'\Gamma_{\approx} + U_1' \Sigma_{\approx}]) 
$$
$$
+ [U_1'\Gamma_1+U_1'\Sigma_1] (-3A'\Gamma_{\lambda} - 3A'\Gamma_{\rho} + (A'\Gamma_1)(3U_1'\Gamma_1 + 2U_1'\Sigma_1)) + (A'\Gamma_1)(M'\Sigma_1)[2U_1'\Gamma_M+U_1'\Sigma_M]
$$
We use the following Ward identities for individual graphs:
$$
(\Gamma_1)_{\mu}(p,0)+\frac{\partial\Sigma_1(p)}{\partial p^{\mu}}=0,\quad 2(\Gamma_M)_{\mu}(p,0)+\frac{\partial\Sigma_M(p)}{\partial p^{\mu}}=0,
$$
$$
2(\Gamma_{\lambda})_{\mu}(p,0) + (\Gamma_{\times})_{\mu}(p,0) +  \frac{\partial\Sigma_{\times}(p)}{\partial p^{\mu}}=0,\quad 2(\Gamma_{\rho})_{\mu}(p,0) + (\Gamma_{\approx})_{\mu}(p,0) +  \frac{\partial\Sigma_{\approx}(p)}{\partial p^{\mu}}=0
$$
Each identity contains one lepton self-energy graph and all vertexlike graphs obtained by inserting an external photon into a line on the main path. Each coefficient equals $1$, but some of our coefficients equal $2$, because we are working with undirected Feynman graphs. The identities can be proved by standard methods used in QED\footnote{See \cite{peskin}.}. The corollary is
$$
U_1'\Gamma_1+U_1'\Sigma_1=0,\quad 2U_1'\Gamma_M+U_1'\Sigma_M=0,\quad 2U_1'\Gamma_{\lambda}+U_1'\Gamma_{\times}+U_1'\Sigma_{\times}=0,\quad 2U_1'\Gamma_{\rho}+U_1'\Gamma_{\approx}+U_1'\Sigma_{\approx}=0.
$$
From this it follows that all terms with $U_1'$ are cancelled. Similarly, we can prove that all terms with $L'$ and $B'$ are lifted in the on-shell expression, which implies the equivalence\footnote{provided that all regularization issues are resolved; dimensional regularization can be used in most cases.}. 

\begin{figure}[H]
	\begin{center}
		\includegraphics[width=60mm]{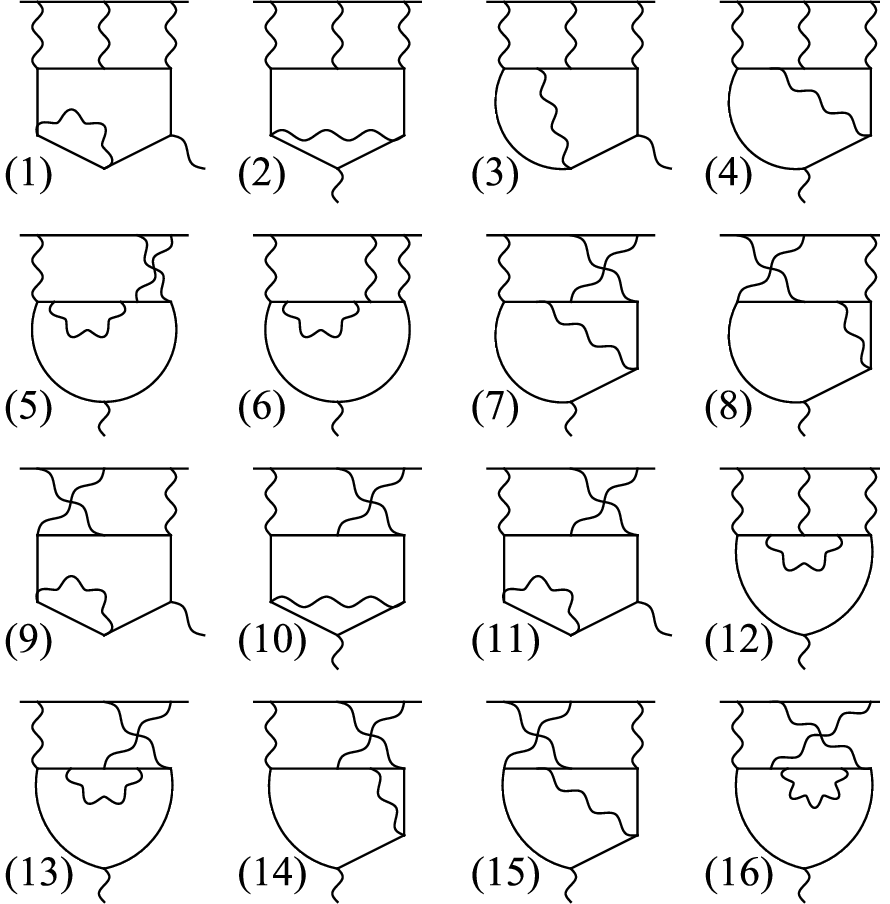}
		\ \ \ \ \ \ 
		\includegraphics[width=52mm]{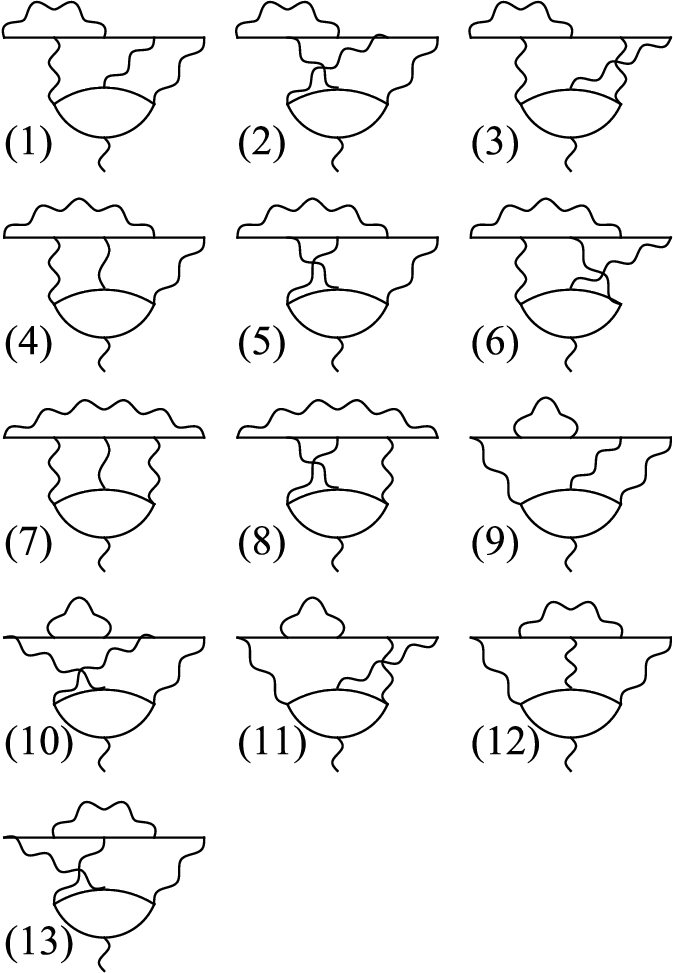}
		\vspace{-3mm}
		\caption{\label{fig_IV_b_c}4-loop gauge-invariant classes IV(b) (left) and IV(c) (right) from ~\cite{kinoshita_8_muon}.}
	\end{center}
\end{figure}

Let us examine examples where the new and old methods work differently. We start with the 4-loop class IV(b) from Fig. \ref{fig_IV_b_c},left. We suppose that the graphs contribute to the muon $g-2$ and contain an electron loop.

The difference is that the old method yields additional subtractions in graphs 2,10 because each of them has a vertexlike subgraph joining the external photon and not lying on the main path\footnote{These subgraphs correspond to fictitious IR divergences, but the subtraction was necessary in \cite{volkov_2015} to make the procedure equivalent to on-shell renormalization.}. Also, the operators $U_2$ are used everywhere instead of $U$. The contribution of IV(b) is equal to
\begin{figure}[H]
	\begin{center}
		\includegraphics[width=85mm]{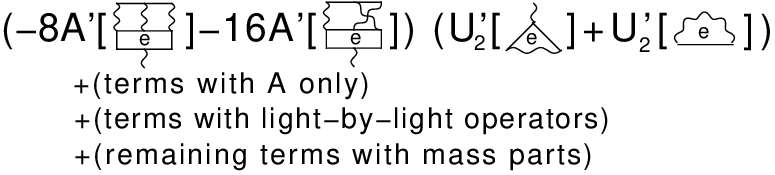}
	\end{center}
\end{figure}
The first term is cancelled because its second multiplier equals $0$ due to the Ward identity; the light-by-light term is similar and is cancelled for the same reason. The expression for the on-shell renormalization has the same form, but the occurrences of $U_2'$ are replaced with $L'$, $B'$; its terms already considered are cancelled due to the same reason. The remaining terms are identical.

Consider the class IV(c) from Fig. \ref{fig_IV_b_c},right. In this case, we also assume that the graphs contribute to the muon $g-2$ and have an electron loop. Graphs 7 and 8 give terms with $U_3$. The total contribution of these graphs is
\begin{figure}[H]
	\begin{center}
		\includegraphics[width=75mm]{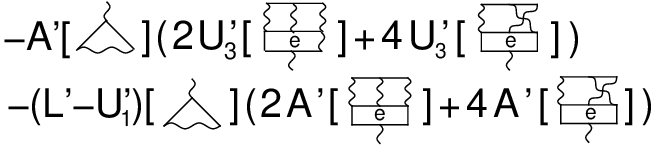}
	\end{center}
\end{figure}
We ignore the terms with $U_0'$ and $M'$ in this investigation, because they do not cause any problems. To show that the first term equals $0$, we need the Ward identities of the form $\Gamma_{\mu}(p,0)=0$ for single graphs. It holds for sets of vertexlike graphs that have the external photon on a \emph{lepton loop} that are closed under the movement of it along the loop. The validity can be proved in the usual way. Since $U_3$ preserves this kind of Ward identities, its contributions are cancelled. The remaining graphs yield
\begin{figure}[H]
	\begin{center}
		\includegraphics[width=75mm]{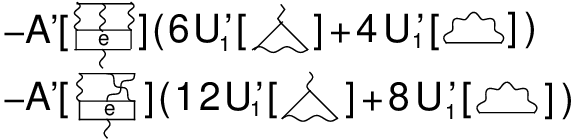}
	\end{center}
\end{figure}
With $U_1'\Gamma_1=-U_1'\Sigma_1$, we get that all terms containing $U_1'$ are cancelled after summation. The total contribution obtained by the in-place on-shell renormalization is
\begin{figure}[H]
	\begin{center}
		\includegraphics[width=85mm]{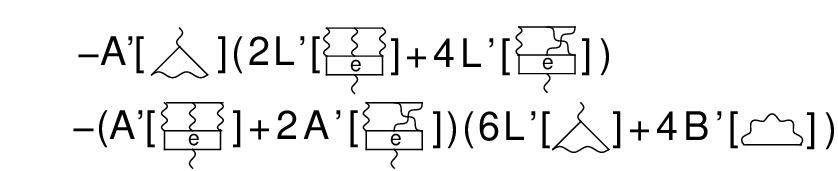}
	\end{center}
\end{figure}
The difference between our and on-shell contributions equals
\begin{figure}[H]
	\begin{center}
		\includegraphics[width=85mm]{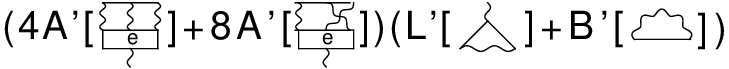}
	\end{center}
\end{figure}
It equals $0$, because $L'\Gamma_1=-B'\Sigma_1$.

Let us summarize. To prove the equivalence, we need two kinds of the ``Ward identities'' for single graphs:
\begin{itemize}
\item $\Gamma_{\mu}(p,0)=-\partial \Sigma(p)/\partial p^{\mu}$; the set includes one lepton self-energy graph and all vertexlike graphs obtained from it by inserting an external photon into a line on the main path.
\item $\Gamma_{\mu}(p,0)=0$; the set consists of vertexlike graphs having the external photon on a lepton loop; it is closed under the movement of the external photon along the loop.
\end{itemize}
Graphs in the sets may contain special vertices. See also examples of the use of the Ward identities for individual graphs in Section IV.H of \cite{volkov_gpu}.

\subsection{Complete proof}\label{subsec_on_shell_proof}

Suppose we have a class $\cckk$ of \emph{1-particle irreducible} vertexlike Feynman graphs that is closed under the movement of internal photons along lepton loops and paths, but without jumping over the external photon. We prove that its contribution obtained with our method is equal to the contribution obtained with the in-place on-shell renormalization. Since the ideas used here are similar to those widely used in QED, the explanation is just a framework.

Let us give some definitions.

Suppose we have one term $X$ of the expression with linear operators constructed for some graph $G$ of this class. The subgraphs to which the operators in $X$ are applied form a rooted tree; in this tree, $G$ is a root, and $G'$ is a descendant of $G''$ if $G'\subset G''$. The contribution of $X$ can be expressed with $A'$, $U_j'$, $L'$, $L'-U_j'$, $M'$, $B'$ applied to some graphs with special vertices. This expression can have several terms $Y_1,\ldots,Y_n$, because $U_j\Sigma$ is split into $U_j'$-part and $M'$-part. Each term $Y_j$ can be represented as a \emph{layer tree}. A \emph{layer} is a Feynman graph obtained from one graph to which an operator in $X$ is applied by schrinking all its childs to special vertices (corresponding to operators in $Y$). A layer tree is a rooted tree, each node of it is $(l,O,r)$, where $l$ is a layer; $O$ is one of the operators $A'$, $U_j'$, $L'$, $L'-U_j'$, $M'$, $B'$; $r$ is the reference to the corresponding vertex of the parent (it does not exist for the root). The root of the layer tree corresponds to $G$ in $X$. Vertices and inputs of one vertex that have the same type are enumerated\footnote{The enumeration is needed to avoid symmetry problems with 4-photon vertices. If we have an enumeration, the coefficients should be properly accounted for; this can be done with a standard technique; we do not do this in our consideration. An  examination of the coefficients is only necessary for symmetry issues; no reduction of similar terms takes place in our consideration.}. The contribution of the layer tree (without coefficient) is the product of the node values\footnote{Strictly speaking, tensor contractions are needed when we have $U_0'$ layers, because for simplicity we decided to use one label for all kinds of special vertices that can be generated. For example, in the definition (\ref{eq_u0_pse_def}), (\ref{eq_u0s_pse_def}) for photon self-energy graphs, $U_0'$ is a vector.}, where the value of the node $(l,O,r)$ is $O\Gamma_l$, where $\Gamma_l$ is the Feynman amplitude corresponding to the layer $l$. For simplicity, we will sometimes interchange the layer tree nodes with the layers.

\begin{figure}[H]
	\begin{center}
		\includegraphics[width=100mm]{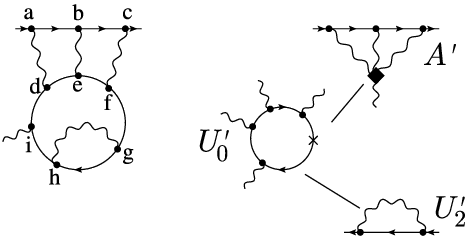}
		\caption{\label{fig_example_layers_4ph}An example of a Feynman graph (left) and a layer tree (right) that is possible for it (with $U_0'$ and $U_2'$ for lepton self-energy graphs)}
	\end{center}
\end{figure}

For example, the term $A_G(U_0)_{defghi} (U_2)_{gh}$ for the graph $G$ from Fig. \ref{fig_example_layers_4ph},left gives the layer tree depicted in Fig. \ref{fig_example_layers_4ph},right.

The following markers are used for vertices on images:
\begin{enumerate}
\item a round dot is an ordinary $\gamma_{\nu}$ vertex;
\item a boxed dot is a mass vertex (which gives $1$);
\item a cross-dot is a $(\slashed{p}-m)$-vertex;
\item a cross in a circle is a vertex coming from photon self energy subgraphs;
\item a diamond dot is a vertex originating from photon-photon scattering subgraphs.
\end{enumerate}

\begin{figure}[H]
	\begin{center}
		\includegraphics[width=140mm]{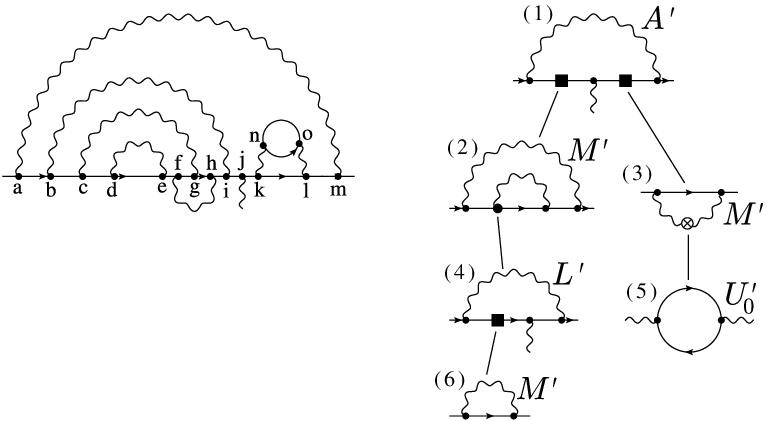}
		\caption{\label{fig_example_layers_m}An example of a Feynman graph (left) and a possible layer tree for it (right) with $M'$-operators}
	\end{center}
\end{figure}

Another example is the term
$$
A_G B_{bcdefghi} L_{cdefg} B_{de} B_{klno} (U_0)_{no}
$$
for $G$ from Fig. \ref{fig_example_layers_m},left, where $M'$ parts are taken for each $B$ (the term is part of the direct on-shell renormalization expression). Its layer tree is depicted in Fig. \ref{fig_example_layers_m},right. The root is layer 1.

The sequence of layer tree nodes $Y_1,\ldots,Y_n$ of vertexlike type is called the \emph{main branch} if the following conditions are satisfied: 
\begin{itemize}
\item $Y_1$ is the root of the layer tree;
\item $Y_{j+1}$ is the child of $Y_j$ corresponding to the vertex of $Y_j$ joining its external photon;
\item $Y_n$ has no vertexlike children corresponding to its vertex joining its external photon;
\end{itemize}
the \emph{type} of a layer is its multiset of the external line types.

We say that the node $X$ of a layer tree is \emph{locally on a path}, if the corresponding to $X$ parent's vertex lies on parent's main path; we also assume that the root of the layer tree is locally on a path. If the node is not locally on a path, it may be on a loop or on a special 2-photon or 4-photon vertex of the parent.

The maximal initial segment $Y_1,Y_2,\ldots,Y_k$ of the main branch $Y_1,Y_2,\ldots,Y_n$, all elements of that are locally on a path, is called the \emph{$\infragr$-branch}.

Let us proceed with the proof. We represent the contribution of $\cckk$ obtained by our method as the sum of the layer tree contributions (with coefficients). Each layer tree has one node $X$ on the $\infragr$-branch that has the operator $A'$. We call it the \emph{A-node}. The operators corresponding to vertexlike or lepton self-energy nodes $X$ satisfy the following rules:
\begin{itemize}
\item if $X$ is the A-node, then the corresponding operator is $A'$;
\item if $X$ is the root but not A-node, then the operator is $L'-U'_1$;
\item if $X$ is an ascendant of the A-node but not the root, then the operator is $L$;
\item if $X$ is a descendant of the A-node and belongs to the $\infragr$-branch, then the operator is $U_{\ww}$, where (\ref{eq_w}) is satisfied for graphs $G$ in $\cckk$;
\item if $X$ is not on the $\infragr$-branch and all ascendants of $X$ and $X$ itself are locally on a path, then the operator is $U_1'$ or $M'$;
\item if $X$ is not on the $\infragr$-branch and at least one ascendant of $X$ or $X$ itself is not locally on a path, then the operator is $U_2'$ or $M'$.
\end{itemize}

Let us perform the following operations with this layer tree sum:
\begin{enumerate}
\item \emph{Remove all terms containing layers of photon-photon scattering type}. In fact, they are cancelled by the identities for individual photon-photon scattering graphs of the form
$$
\Gamma_{\mu_1 \mu_2 \mu_3 \mu_4}(0,0,0,0)=0,
$$
where $\Gamma$ is the corresponding Feynman amplitude, and the external momenta are in the parentheses. The identity is satisfied for a given set of graphs, if the set is closed under the movement of external photons along lepton loops, and \emph{every graph in this set contains at least one external photon line joining a lepton loop} (we consider graphs with special vertices; thus an external photon can fall into a 4-photon vertex). The identities can be proved with a standard technique that is used in QED\footnote{See \cite{peskin}.}. Let us consider the elimination in detail. We say that the layer trees $T_1$ and $T_2$ are equivalent if one is obtained from the other by the movement of external photons along lepton loops in some layers of photon-photon scattering types \emph{which have no descendants of photon-photon scattering type} (the absence of the descendants guarantees that all moved photons are on lepton loops\footnote{They cannot fall to 2-photon vertices due to 1-particle irreducibility.}); we suppose that the structure of the descendants is preserved when a photon is shifted; the children referring to the vertex joining the moved photon change their reference to the new photon location on a lepton loop. It is easy to see that if $T_1$ is obtained from a Feynman graph of $\cckk$, this is also true for $T_2$. Therefore, all terms are split into equivalence classes. In the contribution of each of these classes, the identities mentioned above are factorized. This leaves only the classes that have no photon-photon scattering layers at all. Note that the movement of photons (both external and internal) along lepton loops does not change the 1-particle irreducibility.
\item \emph{Remove all terms containing vertexlike layers with a non-$A'$ operator and the external photon on a loop}\footnote{We consider here only the internal structure of the layer, not its reference to its parent.}. The operator is $L'$, $U_1'$, $U_2'$, $U_3'$. All these operators yield $0$ for the Feynman amplitudes $\Gamma_{\mu}(p,q)$ satisfying $\Gamma_{\mu}(p,0)=0$. Thus all these terms are cancelled by the identities of this form mentioned in Section \ref{subsec_equivalence_examples}. This can be proved in a similar way as for the case with photon-photon scattering layers. After this step, all occurrences of $U_3'$ are removed, since at least the $U_3'$ corresponding to the end of the $\infragr$-branch is applied to a layer with the external photon on a lepton loop.
\item \emph{Represent the layer trees as trees of M-blocks}. An \emph{M-block} includes a node (vertexlike or lepton self-energy-like) with an operator $A'$, $L'$, $U_j'$ ($j=1,2$) of the layer tree, all its children with operators $M'$, $U_0'$, their children with these operators and so on\footnote{We do not use here that $M'$ properly extracts the mass part, but it is used for the proof of gauge invariance \cite{cvitanovic_gauge}.}. For example, there are two M-blocks in \ref{fig_example_layers_m},right: $\{1,2,3,5\}$ and $\{4,6\}$; the latter is a child of the former and refers to the vertex of layer 2 to which layer 4 refers. M-blocks can also be represented as layers. The layer is obtained from the corresponding subgraph by shrinking the subgraphs corresponding to its children (in the tree of M-blocks) to special vertices (in a sense, it is obtained by gluing the layer tree layers belonging to the M-block). An M-block can also be viewed as a tree of nodes. Let us define the \emph{main vertex} of the M-block; if the layer has its external photon on its main path, the main vertex is layer's vertex joining its external photon; otherwise, it is the \emph{last vertex of the main path} of the layer.
\item \emph{Single out the $\ssbr$-branch of M-blocks in each layer tree}. We define the $\ssbr$-branch of M-blocks $Y_1',\ldots,Y_n'$ in a tree of M-blocks by the following conditions:
\begin{itemize}
\item $Y_1'$ is the root of the M-block tree;
\item $Y'_j$, $j\geq 2$, is a child of $Y'_{j-1}$ and refers to the main vertex of $Y'_{j-1}$;
\item there are no children of $Y_n'$ that refer to the main vertex of $Y_n'$.
\end{itemize}
An initial segment $Y_1',\ldots,Y_k'$ of the $\ssbr$-branch corresponds to the $\infragr$-branch $Y_1,\ldots,Y_k$. Due to the 1-particle irreducibility of each subgraph to which an operator is applied, all elements of the $\ssbr$-branch are of vertexlike type.
\item \emph{For each M-block take a one-to-one correspondence between the layer vertices (except the main vertex) and its $\Sigma$-chains lying on the same lepton path part or loop.} A \emph{$\Sigma$-chain} is a peace of a lepton path or loop, which ends are ordinary $\gamma_{\mu}$ vertices, but all intermediate vertices are of $(\slashed{p}-m)$ type (the situation when there are no intermediate vertices is also possible). We require that the $\Sigma$-chain lies on the same side of the M-block external photon as the corresponding vertex (if it is applicable). We should take this correspondence \emph{deterministically} (it should depend only on the topology of the $\Sigma$-chains and external lines). For example, for each $v$ on a lepton loop, we can take the outgoing from $v$ chain; for $v$ on the main path we can take the $\Sigma$-chain joining $v$ and having another end closer to the main vertex of the M-block. 
\item \emph{Remove all terms containing M-blocks outside the $\ssbr$-branch}. Let us explain how they are cancelled. An M-block is called \emph{factorizable} if it has at least one child that does not refer to its main vertex. We say that the M-block trees $\mathcal{T}_1$ and $\mathcal{T}_2$ are equivalent if one is obtained from the other by a sequence of the following transformations:
\begin{itemize}
\item suppose $Y'$ is a factorizable M-block that \emph{has no factorizable children}, $v$ is an ordinary $\gamma_{\mu}$-vertex in $Y'$, the main vertex of $Y'$ is not equal $v$, the $\Sigma$-chain $\sigma$ corresponds to $v$, the child M-block referring to $v$ exists and is denoted by $Y''$; the transformation is: remove the external photon in the \emph{root layer} of $Y''$, move the reference of $Y''$ to the new $(\slashed{p}-m)$ vertex inserted at the beginning of $\sigma$; the operator applied is $U_j'$, $j=1,2$ and remains the same after the transformation; the structure of the descendants and their references are preserved, \emph{the reference to the main vertex of $Y''$ is moved to the main vertex of the transformed M-block} (there are no references to non-main vertices, since $Y'$ has no factorizable children).
\item if under the same conditions there are no children referring to $v$, and the intermediate vertexes of $\sigma$ are $w_1,\ldots ,w_r$, $r\geq 1$ (ordered along the line direction), $Y''$ is the child M-block referring to $w_1$, then insert an external photon into an arbitrary line on the main path of the \emph{root layer} of $Y''$, remove $w_1$ and move the reference of $Y''$ to $v$. the applied operator is $U_j'$, $j=1,2$ and remains the same after the transformation; the structure of the descendants and their references are preserved (as in the previous case).
\end{itemize}
These transformations are inverse to each other. They are illustrated in Fig. \ref{fig_l_to_b_transform}; here $v$ is the round dot, $\sigma$ starts from $v$, $T_0$ is a subtree referring to $v$ in the layer tree, the subtrees $T_1,\ldots,T_n$ refer to the intermediate vertexes of $\sigma$; on the right side, $T_0$ is replaced by $T_0'$ (by removing the external photon in the root layer) and is connected to the begin of $\sigma$. 
\begin{figure}[H]
	\begin{center}
		\includegraphics[width=140mm]{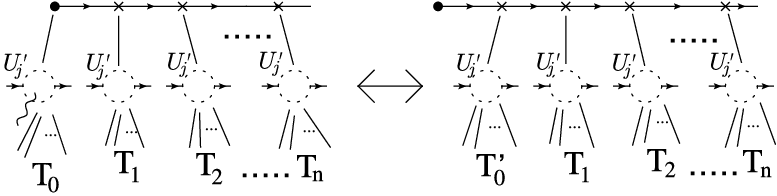}
		\caption{\label{fig_l_to_b_transform}Transformations leading to the cancellation of all terms with M-blocks outside the $\ssbr$-branch.}
	\end{center}
\end{figure}
This equivalence relation splits the terms into equivalence classes. If a term belongs to the expression, all equivalent terms also belong to the expression; it is important here that these transformations do not change the 1-particle irreducibility of the whole graph as well as of the subgraphs to which the operators are applied\footnote{This can be proved using three facts: 1) a Feynman graph and all its subgraphs to which the operators are applied are 1-particle irreducible if and only if each layer of the corresponding layer tree is 1-particle irreducible; 2) adding and removing $(\slashed{p}-m)$ vertices in internal lines does not change the 1-particle irreducibility; 3) inserting the external photon into a lepton line (or removing it) does not change the 1-particle irreducibility.}; the absence of odd lepton loops in the whole graph is also preserved. In the sum of contributions belonging to one equivalence class, the multipliers of the form $U'_j \Gamma + U'_j \Sigma$ (as mentioned in Section \ref{subsec_equivalence_examples}) are factorized. If the number of these multipliers is at least 1, the contribution equals $0$. 

The presence of M-blocks outside the $\ssbr$-branch means that there is at least one factorizable M-block in the $\ssbr$-branch. If it has factorizable children, take one of them; and repeat the operation until it has no factorizable children. So there is at least one factorizable M-block that has no factorizable children. The cancellation of these terms follows from this. 

Note: On the one hand, the correspondence between $v$ and $\sigma$ may cross the boundaries of the layers constituting the M-block; on the other hand, the procedure of inserting and removing the external photon is always performed in the root layer of the M-block, and the structure of $\Sigma$-chains is ignored in this layer (the main path may contain special vertices); but when we move the reference, we use the main vertex of the whole M-block. 
\item \emph{Combine the terms with different placements of the A-nodes.} After the previous step, the contribution can be written as a sum of terms like
\begin{equation}\label{eq_term_after}
(-1)^{n-1}(O_1'\Gamma_1)(O_2'\Gamma_2)\ldots (O_n'\Gamma_n),
\end{equation}
where $\Gamma_1,\ldots,\Gamma_n$ are the vertexlike Feynman amplitudes (after the appropriate application of $-M'$ and $-U_0'$ for subgraphs) corresponding to the $\ssbr$-branch elements $Y'_1,\ldots,Y_n'$, the corresponding linear operators are $O'_1,\ldots,O'_n$. 

The Feynman graph $G$ for which the expression is constructed is uniquely determined from the layers $Y_1',\ldots,Y_n'$ by a sequential insertion of the layers into a point; if $Y_1',\ldots,Y'_{j-1}$ have their external photons on their main paths, then $Y_j'$ is inserted into the vertex joining the external photon; otherwise, it is inserted into the end of the main path.

The term (\ref{eq_term_after}) corresponding to $Y_1',\ldots,Y_n'$, $O_1',\ldots,O_n'$ exists in the expression for $\cckk$ if the following conditions are satisfied:
\begin{itemize}
\item $Y_j'$ are correct vertexlike M-blocks (taking into account their internal layer structure); 
\item $G\in\cckk$, where $G$ is reconstructed by the procedure described above;
\item the vector $(O_1',\ldots,O_n')$ is of the form $(A',U_1',\ldots,U_1')$ or $(L'-U_1',L',\ldots,L',A',U_1',\ldots,U_1')$, where $x,\ldots,x$ may have $0$ elements;
\item for all $i$ such that $O_i'\neq A'$ the layer $Y_i'$ has its external photon on its main path.
\end{itemize}

Any permutation of $Y_1',\ldots,Y_n'$ does not change the validity of $G\in\cckk$ for the reconstructed graph $G$\footnote{To prove this, consider two cases: 1) each $Y_j'$ has its external photon on its main path (in this case the inability to jump over the external photon should be taken into account); 2) at least one $Y_j'$ has its external photon on a lepton loop (in this case the external photon is on a lepton loop).}. Thus, the sum can be represented as a sum of terms (\ref{eq_term_after}) with the same conditions, but with vectors $(O_1',\ldots,O_n')$ of the form $(A',U_1',\ldots,U_1')$ or $(A',L',\ldots,L',L'-U_1',U_1',\ldots,U_1')$.

We can prove by induction that
$$
(U_1',U_1',U_1',\ldots,U_1',U_1')
$$
$$
+(L-U_1',U_1',U_1',\ldots,U_1',U_1')
$$
$$
+(L',L'-U_1',U_1',\ldots,U_1',U_1')
$$
$$
+\ldots
$$
$$
+(L',L',L',\ldots,L',L'-U_1')
$$
$$
=(L',L',L',\ldots,L',L').
$$

Thus, we arrive at the sum of terms
\begin{equation}\label{eq_term_final}
(-1)^{n-1}(A'\Gamma_1)(L'\Gamma_2)\ldots (L'\Gamma_n)
\end{equation}
with the same conditions.
\end{enumerate}

With the contribution of $\cckk$ obtained by the in-place on-shell renormalization\footnote{Note that the definition of the contribution of the gauge-invariant class in \cite{cvitanovic_gauge} already corresponds our final representation; however, it looks like it becomes cumbersome if graphs may have lepton loops.} we perform the same operations, but without the last step and using the identities for $L'$ and $B'$. These operations result in the same sum of terms (\ref{eq_term_final}) with the same conditions.

\section{Computational results}\label{sec_computation}

The new method was tested by numerical calculations. The contributions of 25 gauge-invariant classes contributing to $A_1^{(6)}$, $A_2^{(6)}$, $A_3^{(6)}$ for the muon were presented in \cite{volkov_ffk2021}. Here we present all 4-loop results for the electron and the muon for all gauge-invariant classes (with electron, muon and tau-lepton loops). They are summarized in Table \ref{table_4loop}. We used the values
$$
m_{\mu}/m_e=206.76828103,\quad m_{\tau}/m_{\mu}=16.81665.
$$
in our calculations. The uncertainty of the masses was not taken into account. The aim of the calculation was to verify the method, not to obtain precise physical results (the mass uncertainty does not play a significant role at this level of precision). Some of the previously known values for comparison that are given in the tables were calculated with other values of the mass ratios (and the uncertainties may be inaccurate). The Monte Carlo integration was performed on the ITP/TTP KIT computing cluster with the help of the GPU NVidia A100 and took about three GPU-weeks; a method similar to that described in \cite{volkov_prd, volkov_gpu} was used\footnote{The direct usage of the methods from \cite{volkov_prd, volkov_gpu} is not possible for Feynman graphs having lepton loops. A modification was made; it will be explained in the further papers.}. 

\begin{table}[H]
\caption{\label{table_4loop}The calculated total 4-loop contributions (for the electron and the muon) and their comparison with previously known values; the values for comparison are from the first reference in the lists.}
\begin{ruledtabular}
\begin{tabular}{llll}
Contribution & Our value & Value for comparison & Ref. \\
\colrule 
$A_1^{(8)}$ & -1.9118(41) & -1.91224576 & \cite{laporta_8}; \cite{kinoshita_8_last, smirnov_amm, rappl} \\
$A_2^{(8)}(m_e/m_{\mu})$ & 0.000924(11) & 0.0009141970703(372) & \cite{kurz_8_heavy}; \cite{kinoshita_10_first} \\
$A_2^{(8)}(m_e/m_{\tau})$ & 0.00000710(60) & 0.00000742924(118) & \cite{kurz_8_heavy}; \cite{kinoshita_10_first} \\
$A_3^{(8)}(m_e/m_{\mu},m_e/m_{\tau})$ & 0.000000745(24) & 0.00000074687(28) & \cite{kurz_8_heavy}; \cite{kinoshita_10_first} \\
$A_2^{(8)}(m_{\mu}/m_e)$ & 132.673(84) & 132.6852(60) & \cite{kinoshita_8_muon}; \cite{kurz_8_light} \\
$A_2^{(8)}(m_{\mu}/m_{\tau})$ & 0.04252(11) & 0.0424941(53) & \cite{kurz_8_heavy}; \cite{kinoshita_8_muon} \\
$A_3^{(8)}(m_{\mu}/m_e,m_{\mu}/m_{\tau})$ & 0.0622(33) & 0.062722(10) & \cite{kurz_8_light}; \cite{kinoshita_8_muon}
\end{tabular}
\end{ruledtabular}
\end{table}

\begin{figure}[H]
	\begin{center}
		\includegraphics[width=180mm]{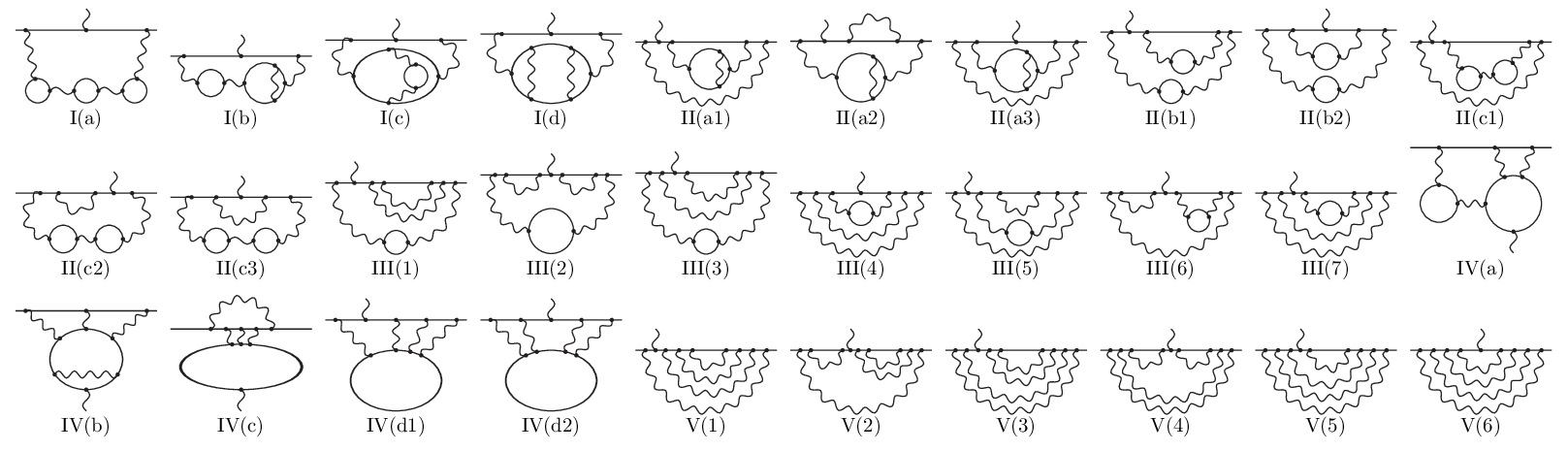}
		\caption{\label{fig_a1_4loops}Gauge-invariant classes contributing to $A_1^{(8)}$. Each class is obtained from the corresponding picture by moving the internal photon ends along the lepton paths and loops, but without jumping over the external photon.}
	\end{center}
\end{figure}

\begin{figure}[H]
	\begin{center}
		\includegraphics[width=180mm]{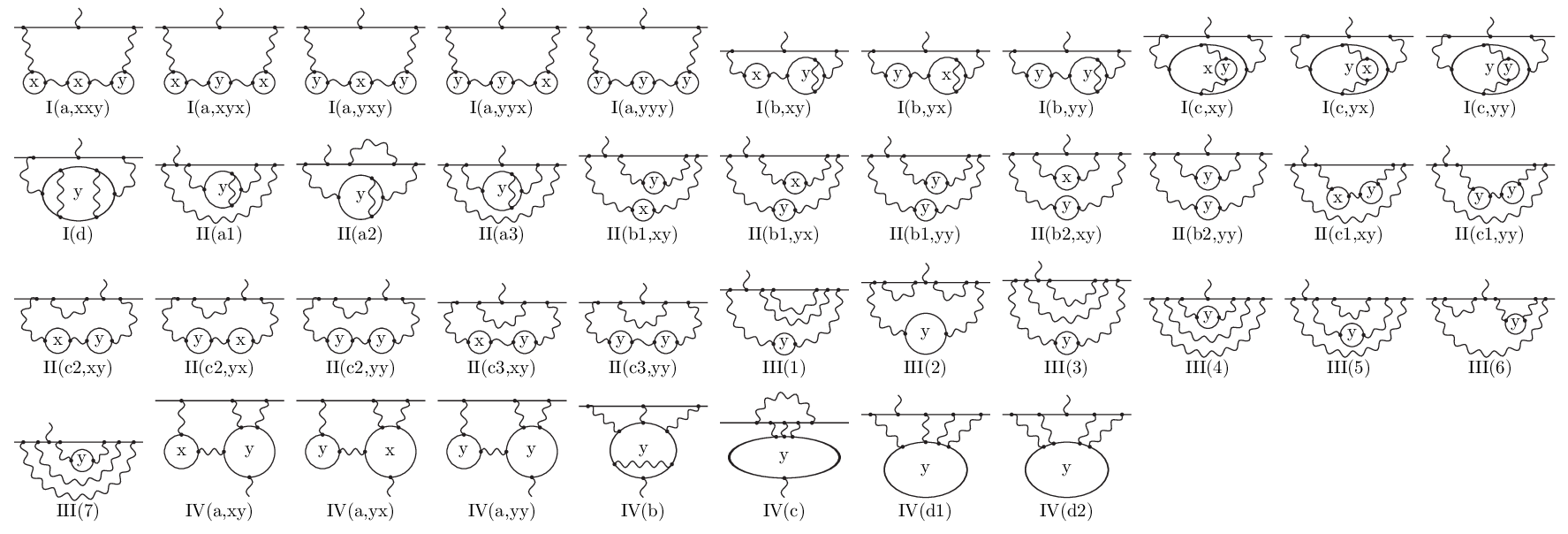}
		\caption{\label{fig_a2_4loops}Gauge-invariant classes contributing to $A_2^{(8)}$ for particle $x$. Each class is obtained from the corresponding image by moving the internal photon ends along the lepton paths and loops, but without jumping over the external photon. $x$ and $y$ may be arbitrary leptons.}
	\end{center}
\end{figure}

\begin{figure}[H]
	\begin{center}
		\includegraphics[width=180mm]{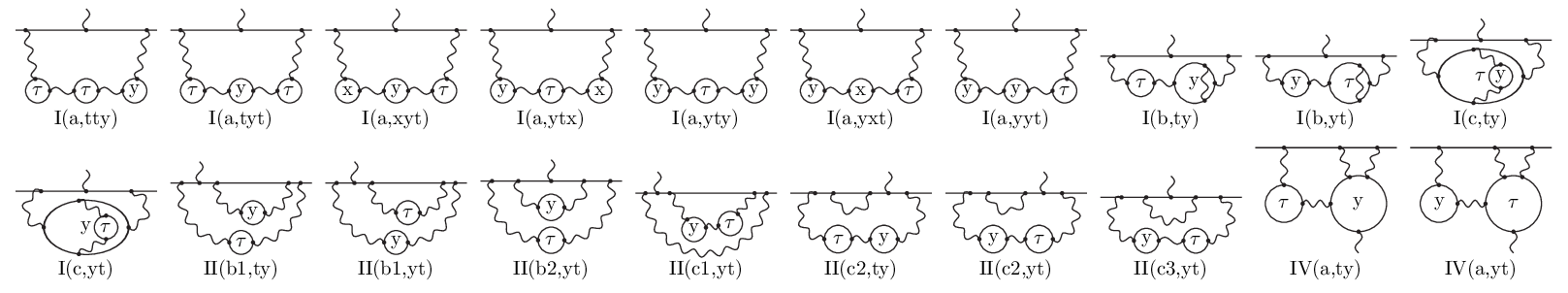}
		\caption{\label{fig_a3_4loops}Gauge-invariant classes contributing to $A_3^{(8)}$ for particle $x$. Each class is obtained from the corresponding image by moving the internal photon ends along the lepton paths and loops, but without jumping over the external photon. $x$ and $y$ are electron and muon (or vice versa), $\tau$ is the tau-lepton.}
	\end{center}
\end{figure}

Table \ref{table_a1} presents the computed contributions of the gauge-invariant classes from Fig. \ref{fig_a1_4loops} to $A_1^{(8)}$ and a comparison with S. Laporta's results \cite{laporta_8}. Table \ref{table_a2} contains all the contributions of the classes from Fig. \ref{fig_a2_4loops} for different particles $x$ and $y$. Since the contribution does not depend on the line directions, all graphs on the images are undirected. Also, permutting the elements on a photon does not change the contribution, but we calculated these contributions separately, except $A_2^{(8)}[\text{II(c2,xy)}]$, $A_2^{(8)}[\text{II(c2,yx)}]$ and $A_3^{(8)}[\text{II(c2,yt)}]$, $A_3^{(8)}[\text{II(c2,ty)}]$. Tables \ref{table_a2_e_comparison} and \ref{table_a2_mu_comparison} contain a comparison with known values for $A_2^{(8)}[\text{electron}]$ and $A_2^{(8)}[\text{muon}]$, respectively. Table \ref{table_a3} contains the contributions of the classes from Fig. \ref{fig_a3_4loops} to $A_3^{(8)}$ for electron and muon, Table \ref{table_a3_comparison} contains their comparison with known results.

Throughout the calculations, the operators
$$
(U_j\Gamma)_{\mu}(p,q)=a(-m^2)\gamma_{\mu},\quad (U_j\Sigma)(p)=s(-m^2)(\slashed{p}-m) + r(m^2) + s(m^2)m
$$
are used for $j=1,2,3$ and all leptons, where (\ref{eq_vertexlike}) and (\ref{eq_sigma}) are satisfied; therefore, $U$-subtractions are performed at space-like momenta ($p^2<0$); this is a general observation: the subtractions at space-like points make the oscillations slightly smaller and the convergence of the Monte Carlo integration slightly faster \cite{volkov_acat_2021}.

For photon self-energy subgraphs, the replacement
$$
\Pi_{\mu\nu}(p^2)\rightarrow p^2 g_{\mu\nu}[h_2(0)-h_2(p^2)],
$$
where (\ref{eq_pi_def}) is implied, was used instead of subtractions. It makes the integrand simpler and its evaluation faster. This approach was used earlier in many calculations; see, for example, \cite{kinoshita_infrared}. Known lower-order analytical expressions for the polarization operator were not used; the reason is the attempt to make everything simple for realization; in addition, the usage of known analytical formulas can improve the calculation speed only for those contributions that already require not too large numbers of the Monte Carlo samples to reach the needed accuracy; this would not affect the whole calculation speed significantly. Moreover, if the number of self-energy subgraphs is small, the ``direct'' realization does not lead to a significant increase in the complexity; this significantly affects only the number of the integration variables (which is not a problem for the Monte Carlo integration, provided that a good probability density function has been chosen).

\begin{table}[H]
\caption{\label{table_a1}Computed contributions of the classes from Fig. \ref{fig_a1_4loops} to $A_1^{(8)}$ and their comparison with the results of \cite{laporta_8}.}
\begin{ruledtabular}
\begin{tabular}{lllllll}
Class & Value & Value from \cite{laporta_8} & \ \ \ &
Class & Value & Value from \cite{laporta_8} \\
\colrule 
I(a) & 0.00087614(64) & 0.0008768659 &  & III(7) & -0.012485(63) & - \\
I(b) & 0.0153300(39) & 0.0153252829 & & IV(a) & 0.598864(69) & 0.598842072 \\
I(c) & 0.0111324(13) & 0.0111309140 & & IV(b) & 0.82236(20) & 0.8222844858 \\
I(d) & 0.049544(60) & 0.0495132026 & & IV(c) & -1.13789(52) & -1.1388228765 \\
II(a1) & -0.255082(24) & - & & IV(d1) & -0.87251(21) & -0.872657392 \\
II(a2) & -0.317402(29) & - & & IV(d2) & -0.117978(94) & -0.1179498688 \\
II(a3) & 0.151986(16) & 0.1519895997 & & V(1) & -1.9723(12) & -1.9710756168 \\
II(b1) & -0.0341771(30) & -0.034179376 & & V(2) & -0.6227(20) & -0.6219210635 \\
II(b2) & 0.0065025(10) & 0.0065041484 & & V(3) & -0.1412(21) & -0.1424873798 \\
II(c1) & -0.0376103(54) & - & & V(4) & 1.0865(15) & 1.0866983948 \\
II(c2) & -0.0536815(84) & - & & V(5) & -1.0397(19) & -1.04054241 \\
II(c3) & 0.0178533(40) & 0.0178536865 & & V(6) & 0.51153(71) & 0.512462048 \\
III(1) & 0.128258(83) & - & & II(a1)+II(a2) & -0.572485(38) & -0.5724718621 \\
III(2) & 0.242948(42) & - & & II(c1)+II(c2) & -0.091292(10) & -0.09130584 \\
III(3) & -0.04388(10) & - & & III(1)+III(5) & 0.69078(12) & 0.6904483476 \\
III(4) & 0.374317(49) & 0.3743579348 & & III(2)+III(6) & 0.409178(76) & 0.4092170285 \\
III(5) & 0.562521(92) & - & &  III(3)+III(7) & -0.05637(12) & -0.0563360902 \\
III(6) & 0.166229(64) & - & & 
\end{tabular}
\end{ruledtabular}
\end{table}

\begin{table}[H]
\caption{\label{table_a2}Calculated contributions of the classes from Fig. \ref{fig_a2_4loops} to $A_2^{(8)}$ for electron and muon.}
\begin{ruledtabular}
\begin{tabular}{lllll}
Class & $\text{Value} \times 10^3,$ $x=e$, $y=\mu$ & $\text{Value} \times 10^5,$ $x=e$, $y=\tau$  & Value, $x=\mu$, $y=e$ & Value, $x=\mu$, $y=\tau$ \\
\colrule 
I(a,xxy) & 0.000168(35) & 0.00003(37) & 0.01854(19) & 0.00002116(31) \\
I(a,xyx) & 0.000077(18) & 0.00008(22) & 0.009378(80) & 0.00001057(16) \\
I(a,yxy) & 0.0000009(22) & -0.000004(13) & 0.16473(32) & 0.000000274(38) \\
I(a,yyx) & 0.0000034(45) & 0.000012(27) & 0.32986(60) & 0.000000612(77) \\
I(a,yyy) & 0.000000006(22) & 0.0000000035(78) & 7.2235(18) & 0.0000000180(35) \\
I(b,xy) & 0.000963(51) & 0.00050(78) & 0.1195(10) & 0.00014580(54) \\
I(b,yx) & 0.000811(79) & 0.0010(11) & 0.33397(53) & 0.00010541(69) \\
I(b,yy) & 0.00000038(20) & 0.000000053(79) & 7.1267(62) & 0.000001950(32) \\
I(c,xy) & 0.004754(57) & 0.00346(41) & 0.16205(25) & 0.00027729(30) \\
I(c,yx) & 0.005936(32) & 0.00361(37) & 0.02156(40) & 0.00040798(35) \\
I(c,yy) & 0.0003543(26) & 0.000127(10) & 1.4372(25) & 0.000053149(67) \\
I(d) & 0.002457(32) & 0.000915(62) & -0.234(15) & 0.0003681(11) \\
II(a1) & -0.041120(60) & -0.02104(22) & -2.3873(57) & -0.0037194(13) \\
II(a2) & -0.07041(11) & -0.04147(57) & -2.4684(73) & -0.0054854(21) \\
II(a3) & 0.025403(57) & 0.01748(23) & 2.0651(80) & 0.0018698(11) \\
II(b1,xy) & -0.017667(63) & -0.01333(56) & -0.37697(57) & -0.00098634(55) \\
II(b1,yx) & -0.027301(55) & -0.02399(68) & -0.31493(55) & -0.00126397(53) \\
II(b1,yy) & -0.00090489(12) & -0.000319911(44) & -6.2229(30) & -0.000136912(20) \\
II(b2,xy) & 0.006692(24) & 0.00684(22) & 0.16083(31) & 0.00033086(24) \\
II(b2,yy) & 0.000133460(43) & 0.000047178(15) & 2.1911(19) & 0.0000204864(67) \\
II(c1,xy) & -0.04082(29) & -0.0306(33) & -0.6564(10) & -0.0022605(23) \\
II(c1,yy) & -0.0009045(88) & -0.000336(36) & -6.0791(24) & -0.00014068(19) \\
II(c2,xy) & -0.04159(30) & -0.0373(30) & -0.4278(10) & -0.0019306(26) \\
II(c2,yx) & -0.04159(30) & -0.0373(30) & -0.4278(10) & -0.0019306(26) \\
II(c2,yy) & -0.002274(22) & -0.000921(84) & -6.7191(37) & -0.00027518(40) \\
II(c3,xy) & 0.03193(20) & 0.0337(19) & 0.36033(65) & 0.0011914(16) \\
II(c3,yy) & 0.0010686(85) & 0.000539(36) & 4.6173(27) & 0.00009622(17) \\
III(1) & -0.0384(11) & -0.0679(70) & 1.7884(74) & 0.000526(12) \\
III(2) & 0.23468(44) & 0.2280(34) & 1.7049(78) & 0.0098734(51) \\
III(3) & 0.0860(23) & 0.073(10) & -2.703(14) & 0.003569(14) \\
III(4) & 0.2010(17) & 0.1436(61) & 4.375(14) & 0.0111379(75) \\
III(5) & 0.2768(11) & 0.2027(75) & 4.400(10) & 0.016024(11) \\
III(6) & 0.05621(29) & 0.0395(15) & 2.565(10) & 0.0035162(48) \\
III(7) & 0.00390(33) & -0.0013(16) & -1.3382(82) & 0.0005628(56) \\
IV(a,xy) & 0.1540(17) & 0.054(16) & 2.6961(53) & 0.012611(17) \\
IV(a,yx) & 0.4639(13) & 0.364(13) & 4.3409(74) & 0.023262(12) \\
IV(a,yy) & 0.01755(10) & 0.00658(34) & 116.802(27) & 0.0026298(27) \\
IV(b) & 0.04174(90) & 0.0118(30) & -0.423(52) & 0.006132(20) \\
IV(c) & -0.1954(50) & -0.114(25) & 2.860(47) & -0.018337(83) \\
IV(d1) & -0.1693(70) & -0.092(38) & -3.5037(79) & -0.015203(55) \\
IV(d2) & -0.0051(45) & -0.000(28) & -0.9187(49) & -0.000555(22)
\end{tabular}
\end{ruledtabular}
\end{table}

\begin{table}[H]
\caption{\label{table_a2_e_comparison}Comparison with previously known values for $A_2^{(8)}$ of the electron ($x=e$); the values for comparison are from the first reference in the lists.}
\begin{ruledtabular}
\begin{tabular}{p{2 cm}lllll}
Class & $\text{Value} \times 10^3,$ $y=\mu$ & Value for comparison & $\text{Value} \times 10^5,$ $y=\tau$  & Value for comparison & Ref. \\
\colrule 
I(a) & 0.000250(40) & 0.0002264 & 0.00012(43) & 0.0000802 & \cite{kurz_8_heavy}; \cite{kinoshita_10_first,muon_8_vacpol_laporta, solovtsova_2019, solovtsova_2022} \\
I(b) & 0.001774(94) & 0.001704139(76) & 0.0015(14) & 0.000602805(26) & \cite{kinoshita_10_first}; \cite{muon_8_vacpol_laporta}  \\
I(c) & 0.011044(66) & 0.0110072(15) & 0.00720(55) & 0.0069819(12) & \cite{kinoshita_10_first} \\
I(d) & 0.002457(32) & 0.0024727  & 0.000915(62) & 0.0008746 & \cite{kurz_8_heavy}; \cite{kinoshita_10_first}  \\
II(a) & -0.08613(14) & -0.0864460(90) & -0.04503(65) & -0.0456480(70) & \cite{kinoshita_10_first}; \cite{muon_8_vacpol_laporta} \\
II(b) & -0.039048(87) & -0.0390003(27) & -0.03075(90) & -0.0303937(42) & \cite{kinoshita_10_first} \\
II(c) & -0.09418(55) & -0.095097(24) & -0.0721(57) & -0.071697(25) & \cite{kinoshita_10_first} \\
III & 0.8203(33) & 0.81715 & 0.617(16) & 0.6059 & \cite{kurz_8_heavy}; \cite{kinoshita_10_first} \\
IV(a) & 0.6354(21) & 0.63579 & 0.425(21) & 0.451 & \cite{kurz_8_heavy}; \cite{kinoshita_10_first} \\
IV(b) & 0.04174(90) & 0.041574 & 0.0118(30) & 0.01471 & \cite{kurz_8_heavy}; \cite{kinoshita_10_first,kataev_8,melnikov_8} \\
IV(c) & -0.1954(50) & -0.19548262120(20) & -0.114(25) & -0.0979 & \cite{kurz_8_heavy}; \cite{kinoshita_10_first} \\
IV(d) & -0.1744(83) & -0.1778(12) & -0.092(47) & -0.0927(13) & \cite{kinoshita_10_first} \\
I(a,yyy) & 0.000000006(22) & 0.000000001033 &  0.0000000035(78) & 0.000000000001293 & \cite{solovtsova_2019}; \cite{solovtsova_2022, muon_8_vacpol_laporta} \\
I(a,yxy) +I(a,yyx) & 0.0000043(51) & 0.0000001346 &  0.000008(30) & 0.0000000003954 & \cite{solovtsova_2019}; \cite{solovtsova_2022, muon_8_vacpol_laporta} \\
I(a,xxy) +I(a,xyx) & 0.000246(40) & 0.0002263 & 0.00011(43) & 0.00008025 & \cite{solovtsova_2019}; \cite{solovtsova_2022, muon_8_vacpol_laporta} \\
I(b,xy) & 0.000963(51) & 0.0010008 & 0.00050(78) & 0.000354 & \cite{muon_8_vacpol_laporta} \\
I(b,yx) & 0.000811(79) & 0.00070428 & 0.0010(11) & 0.000249 & \cite{muon_8_vacpol_laporta} \\
I(b,yy) & 0.00000038(20) & 0.00000017906 & 0.000000053(79) & 0.000000000354 & \cite{muon_8_vacpol_laporta} \\
II(c1,yy) +II(c2,yy) +II(c3,yy) & -0.002110(25) & -0.0021216 & -0.00072(10) & -0.00075 & \cite{muon_8_vacpol_laporta} \\
II(c1,xy) +II(c2,xy) +II(c2,yx) +II(c3,xy) & -0.09207(55) & -0.092981 & -0.0714(57) & -0.0710 & \cite{muon_8_vacpol_laporta} 
\end{tabular}
\end{ruledtabular}
\end{table}

\begin{table}[H]
\caption{\label{table_a2_mu_comparison}Comparison with previously known values for $A_2^{(8)}$ of the muon ($x=\mu$); the values for the comparison are from the first reference in the lists.}
\begin{ruledtabular}
\begin{tabular}{p{2 cm}llllll}
Class & $\text{Value},$ $y=e$ & Value for comparison & Ref. & $\text{Value},$ $y=\tau$  & Value for comparison & Ref. \\
\colrule 
I(a) & 7.7460(19) & 7.745136 & \cite{kurz_8_light}; \cite{kinoshita_8_muon} & 0.00003263(36) & 0.00003242810(20) & \cite{kurz_8_heavy}; \cite{kinoshita_8_muon} \\
I(b) & 7.5802(63) & 7.58201(71) & \cite{kinoshita_8_muon} & 0.00025316(87) & 0.000252 & \cite{kinoshita_8_muon} \\
I(c) & 1.6208(25) & 1.624307(40) & \cite{kinoshita_8_muon} & 0.00073843(47) & 0.000737 & \cite{kinoshita_8_muon} \\
I(d) & -0.234(15) & -0.2303620(50) & \cite{muon_8_baikov_broadhurst}; \cite{kinoshita_8_muon, kurz_8_light} & 0.0003681(11) & 0.0003677960(40) & \cite{kurz_8_heavy}; \cite{kinoshita_8_muon} \\
II(a) & -2.791(12) & -2.77885 & \cite{kurz_8_light}; \cite{kinoshita_8_muon, muon_8_vacpol_laporta} & -0.0073350(26) & -0.0073290(10) & \cite{kinoshita_8_muon} \\
II(b) & -4.5628(37) & -4.55277(30) & \cite{kinoshita_8_muon} & -0.00203588(80) & -0.002036 & \cite{kinoshita_8_muon} \\
II(c) & -9.3325(55) & -9.34180(83) & \cite{kinoshita_8_muon} & -0.0052501(46) & -0.0052460(10) & \cite{kinoshita_8_muon} \\
III & 10.792(28) & 10.7934(27) & \cite{kinoshita_8_muon}; \cite{kurz_8_light} & 0.045209(24) & 0.0452089860(60) & \cite{kurz_8_heavy}; \cite{kinoshita_8_muon} \\
IV(a) & 123.839(29) & 123.78551(44) & \cite{kinoshita_8_muon}; \cite{kurz_8_light} & 0.038504(20) & 0.038519670(30) & \cite{kurz_8_heavy}; \cite{kinoshita_8_muon} \\
IV(b) & -0.423(52) & -0.4170(37) & \cite{kinoshita_8_muon}; \cite{kurz_8_light} & 0.006132(20) & 0.006126610(50) & \cite{kurz_8_heavy}; \cite{kinoshita_8_muon,kataev_8,melnikov_8} \\
IV(c) & 2.860(47) & 2.9072(44) & \cite{kinoshita_8_muon}; \cite{kurz_8_light} & -0.018337(83) & -0.01830100(10) & \cite{kurz_8_heavy}; \cite{kinoshita_8_muon} \\
IV(d) & -4.4225(93) & -4.43243(58) & \cite{kinoshita_8_muon}; \cite{kurz_8_light} & -0.015757(59) & -0.015868(37) & \cite{kinoshita_8_muon} \\
I(a,yyy) & 7.2235(18) & 7.22307640(80) & \cite{muon_8_vacpol_laporta}; \cite{kurz_8_light, aguilar_rafael_8} & 0.000000037(14) & 0.0000000232 & \cite{solovtsova_2019}; \cite{solovtsova_2022} \\
I(b,xy) & 0.1195(10) & 0.1196024600(20) & \cite{muon_8_vacpol_laporta} & - & - & - \\
I(b,yx) & 0.33397(53) & 0.333664680(10) & \cite{muon_8_vacpol_laporta} & - & - & -  \\
I(b,yy) & 7.1267(62) & 7.12800840(20) & \cite{muon_8_vacpol_laporta} & - & - & -  \\
I(c,xy) & 0.16205(25) & 0.161982(11) & \cite{muon_kinoshita_nio_2004} & - & - & -  \\
I(c,yx) & 0.02156(40) & 0.0215830(20) & \cite{muon_kinoshita_nio_2004} & - & - & -  \\
I(c,yy) & 1.4372(25) & 1.440744(16) & \cite{muon_kinoshita_nio_2004} & - & - & -  \\
IV(a,xy) & 2.6961(53) & 2.69(14) & \cite{kurz_8_light} & - & - & -  \\
IV(a,yx) & 4.3409(74) & 4.33(17) & \cite{kurz_8_light} & - & - & -  \\
IV(a,yy) & 116.802(27) & 116.760(20) & \cite{kurz_8_light} & - & - & -  \\
I(a,yxy) +I(a,yyx) & 0.49459(68) & 0.494072030(30) & \cite{muon_8_vacpol_laporta}; \cite{kurz_8_light, aguilar_rafael_8} & 0.000000885(86) & 0.0000008757 & \cite{solovtsova_2019}; \cite{solovtsova_2022} \\
I(a,xxy) +I(a,xyx) & 0.02792(20) & 0.0279883220(70) & \cite{muon_8_vacpol_laporta}; \cite{kurz_8_light, aguilar_rafael_8} & 0.00003173(35) & 0.0000315291 & \cite{solovtsova_2019}; \cite{solovtsova_2022} \\
II(c1,yy) +II(c2,yy) +II(c3,yy) & -8.1808(51) & -8.1895 & \cite{muon_8_vacpol_laporta} & - & - & -  \\
II(c1,xy) +II(c2,xy) +II(c2,yx) +II(c3,xy) & -1.1517(18) & -1.1532 & \cite{muon_8_vacpol_laporta} & - & - & -  \\
II(b1,yy) +II(b2,yy) +II(c1,yy) +II(c2,yy) +II(c3,yy) & -12.2126(63) & -12.212631 & \cite{kurz_8_light} & - & - & -  \\
II(b1,xy) +II(b1,yx) +II(b2,xy) +II(c1,xy) +II(c2,xy) +II(c2,yx) +II(c3,xy) & -1.6828(20) & -1.683165(13) & \cite{kurz_8_light} & - & - & -  
\end{tabular}
\end{ruledtabular}
\end{table}

\begin{table}[H]
\caption{\label{table_a3}Computed contributions of  classes from Fig. \ref{fig_a3_4loops} to $A_3^{(8)}$ for electron and muon.}
\begin{ruledtabular}
\begin{tabular}{llllll}
Class & $\text{Value} \times 10^7,$ $x=e$ & $\text{Value},$ $x=\mu$ & Class & $\text{Value} \times 10^7,$ $x=e$ & $\text{Value},$ $x=\mu$ \\
\colrule 
I(a,tty) & -0.000003(17) & -0.0000012(85) & I(c,yt) & 0.0797(60) & 0.000383(24) \\
I(a,tyt) & 0.0000063(85) & 0.0000038(35) & II(b1,ty) & -0.30252(13) & -0.006434(89) \\
I(a,xyt) & -0.0113(95) & 0.000155(34) & II(b1,yt) & -0.22691(14) & -0.00457(10) \\
I(a,ytx) & -0.0033(93) & 0.000133(30) & II(b2,yt) & 0.070470(61) & 0.002062(38) \\
I(a,yty) & 0.000037(36) & 0.000884(51) & II(c1,yt) & -0.533(33) & -0.00972(23) \\
I(a,yxt) & 0.008(11) & 0.000147(30) & II(c2,yt) & -0.718(63) & -0.00753(20) \\
I(a,yyt) & -0.000022(71) & 0.00184(11) & II(c2,ty) & -0.718(63) & -0.00753(20) \\
I(b,ty) & 0.00025(20) & 0.00069(13) & II(c3,yt) & 0.771(29) & 0.00498(12) \\
I(b,yt) & 0.00001(14) & 0.001944(81) & IV(a,ty) & 5.94(16) & 0.0482(12) \\
I(c,ty) & 0.0965(52) & 0.001389(36) & IV(a,yt) & 2.99(15) & 0.0351(20) \\
\end{tabular}
\end{ruledtabular}
\end{table}

\begin{table}[H]
\caption{\label{table_a3_comparison}Comparison with previously known values for $A_3^{(8)}$ for the electron and the muon; the comparison values are from the first reference in the lists.}
\begin{ruledtabular}
\begin{tabular}{lllllll}
Class & $\text{Value}\times 10^7,$ $x=e$ & Value for comparison & Ref. & $\text{Value},$ $x=\mu$ & Value for comparison & Ref. \\
\colrule 
I(a) & -0.007(17) & 0.0000119956 & \cite{kurz_8_heavy}; \cite{kinoshita_10_first} & 0.00316(14) & 0.003209050(10) & \cite{kurz_8_light}; \cite{kinoshita_8_muon} \\
I(b) & 0.00025(24) & 0.0000140970(10) & \cite{kinoshita_10_first} & 0.00264(16) & 0.002611 & \cite{kinoshita_8_muon} \\
I(c) & 0.1763(79) & 0.172860(21) & \cite{kinoshita_10_first} & 0.001772(43) & 0.001807 & \cite{kinoshita_8_muon} \\
II(b) & -0.45897(20) & -0.458968(17) & \cite{kinoshita_10_first} & -0.00895(14) & -0.0090080(10) & \cite{kinoshita_8_muon} \\
II(c) & -1.20(10) & -1.18969(67) & \cite{kinoshita_10_first} & -0.01980(38) & -0.0196420(20) & \cite{kinoshita_8_muon} \\
IV(a) & 8.93(22) & 8.9432(25) & \cite{kurz_8_heavy}; \cite{kinoshita_10_first} & 0.0834(23) & 0.083747570(90) & \cite{kurz_8_light}; \cite{kinoshita_8_muon} \\
I(a,yty)+I(a,yyt) & - & - & - & 0.00273(12) & 0.00274860(90) & \cite{aguilar_rafael_8}
\end{tabular}
\end{ruledtabular}
\end{table}

\section{Conclusion and discussion}

The development of techniques plays a very important role in theoretical physics. Techniques of a certain kind first appear; then they are refined: they become simpler, less cumbersome, more efficient; and only after that can the technique be extended to broader classes of tasks and lead to a breakthrough. 

This work is a part of the ``refinement'' phase. At present, the described method of divergence subtraction works only for the lepton magnetic moments and QED. However, its formulation is much simpler and compact than that of the previously known methods. Unlike the old method described in \cite{volkov_2015}, the new method has more freedom and does not require useless extra subtractions; but most importantly, the ideas behind the new method appear extendable to broader classes of problems. 

It was supposed that the old method \cite{volkov_2015}  only worked for $A_1^{(n)}$. However, it turned out to be suitable for the computation of $A_2^{(8)}$, $A_3^{(8)}$ as well. This was first discovered numerically and was surprising, but this can be proved: all the extra subtraction terms that the old method has have the operator $L'$ or $L'-U'$ applied to a layer with external photon on a lepton loop; thus, all these occurrences are cancelled, as it was proved in Section \ref{subsec_on_shell_proof} for $U'_3$.

The old method can also lead to the situation where subtracting a fictitious IR divergence produces a real IR divergence (see Discussion in \cite{volkov_2015}). The cancellation of these divergences was conjectured in \cite{volkov_2015}. We have now checked this numerically: for the only 5-loop graph in which this situation occurs, the obtained integral is indeed finite. In addition, the reasoning in Section \ref{subsec_on_shell_proof} looks cumbersome and might contain mistakes; therefore, we checked on a computer that the required combinatorial identities are satisfied, up to 6 loops and 3 particles. However, the fact of divergece cancellation in every single Feynman graph requires a full rigorous proof.

\begin{acknowledgments}
The author would like to thank Gudrun Heinrich, Savely Karshenboim, Andrey Arbuzov, Lidia Kalinovskaya for valuable assistance and Andrey Kataev for help in writing the review section. The computations were performed with the help of the ITP/TTP KIT computing cluster. All images were drawn using AxoDraw and JaxoDraw. The research is supported by the Alexander von Humboldt Foundation. 
\end{acknowledgments}

\bibliography{method_details_2023}

\end{document}